\documentclass[12pt]{iopart}
\usepackage{ifpdf}
\ifpdf
   \usepackage[pdftex]{graphicx}
   \DeclareGraphicsExtensions{.pdf}
   \usepackage{thumbpdf}      
\else
   \usepackage[dvips]{graphicx}
   \DeclareGraphicsExtensions{.eps}
\fi

\usepackage{cite}
\usepackage{amssymb}
\usepackage{amscd}
\usepackage{subfigure}
\usepackage{setstack}
\usepackage{iopams}

\renewcommand{\i}{{\rm i}}

\renewcommand{\text}[1]{{\mbox{#1}}}
\renewenvironment{pmatrix}{\left(\!\!\begin{array}{cc}}{\end{array}\!\!\right)}
\newcommand{\ri}{{ \rm i }}
\newcommand{\re}{{ \rm e }}
\newcommand{\rd}{{ \rm d }}

\newcommand{\abs}[1]{\vert#1\vert}

\newcommand{\be}{\begin{equation}}
\newcommand{\ee}{\end{equation}}

\newcommand{\cn}{{\rm cn}}
\newcommand{\sn}{{\rm sn}}
\newcommand{\dn}{{\rm dn}}

\usepackage{color}
\definecolor{blau}{rgb}{0,0,1}
\definecolor{gruen}{rgb}{0,1,0}
\definecolor{rot}{rgb}{1,0,0}
\definecolor{magenta}{rgb}{1,0,1}

\begin{document}
\jl{1}
\title[Resonance solutions of the NLSE in an open double-well potential]
{Resonance solutions of the nonlinear Schr\"odinger equation in an open double-well potential}
\author{K Rapedius and H J Korsch}

\address{FB Physik, Technische Universit\"at Kaiserslautern, D-67653
Kaiserslautern, Germany}

\ead{korsch@physik.uni-kl.de}

\begin{abstract}
The resonance states and the decay dynamics of the nonlinear Schr\"odinger (or
Gross-Pitaevskii) equation are studied for a simple, however flexible model
system, the double delta-shell potential. This model allows analytical solutions
and provides insight into the influence of the nonlinearity on the decay dynamics.
The bifurcation scenario of the resonance states is discussed, as well as their
dynamical stability properties. A discrete approximation using
a biorthogonal basis is suggested which allows an accurate description even
for only two basis states in terms of a nonlinear, nonhermitian matrix problem.
\end{abstract}


\section{Introduction}
In quantum dynamics metastable states can be conveniently described as resonance
states, i.e.~eigenstates of the Schr\"odinger equation with complex eigenvalues.
Such resonances can be efficiently calculated by complex scaling methods \cite{Mois98} or
matrix truncation techniques for periodic lattices \cite{02wsrep} and found numerous
applications. The recent progress of the physics of Bose-{Einstein} condensates (BEC)
stimulated investigations of the role of resonances for such systems, as for
instance escape from a potential well or decay by loss of condensate particles as it was
realized in a recent experiment with ultracold molecules in a one-dimensional lattice 
\cite{Syas08}. Alternative implementations of open
systems with interactions include experiments with optical wave guide arrays 
\cite{Trom06a,Trom06b,Muss08}.

Here we will consider a (quasi) one-dimensional configuration
described by the nonlinear Schr\"odinger equation (NLSE) or Gross--Pitaevskii equation (GPE)
for the macroscopic condensate wavefunction
\begin{equation}
  \left( -\frac{\hbar^2}{2m} \frac{\partial^2}{\partial x^2} + V(x) +
  g \abs{\psi(x,t)}^2 \right) \psi(x,t) = \ri \hbar \,\frac{\partial \psi(x,t)}{\partial t}
\label{tGPE}
\end{equation}
where the nonlinear parameter $g$ describes the self-interaction.
First applications of complex energy resonance states
extending the complex scaling method to the nonlinear case
investigated the decay of a condensate trapped in the celebrated model potential
$V(x)\sim (x^2-\beta)\re^{-\alpha x^2}$ \cite{Mois04a,Mois05,Schl06a}. However,
the definition of a resonance is somewhat ambiguous in the nonlinear case;
alternative descriptions based on amplitude ratios in the inner and outer
potential region have also been proposed and applied to simple model systems
allowing an analytical treatment \cite{04nls_delta}. One of these models
is the delta-shell potential discussed in section~\ref{sec-single}.

Related studies explore resonance phenomena observed in transport problems
of BECs, for example the transmission though a potential barrier or across
a potential well \cite{Paul05,06nl_transport,Paul07b,08nlLorentz} which
can be explained in terms of an underlying resonance state
\cite{Paul07b,08nlLorentz}.
Complex resonances have also been used to describe a BEC
in accelerated optical lattices by means of nonlinear Wannier-Stark states
\cite{Wimb06,07nlres}. Here the potential in (\ref{tGPE}) is of the form
$V_p(x)+Fx$, where $V_p(x)$ is periodic in space. Such systems can also
be described in terms of Wannier functions localized on the potential minima,
the `sites'.
In a single-band approximation this leads to a discrete nonlinear
Schr\"odinger (or Gross-Pitaevskii) equation, well known as the discrete
self-trapping equation (DST) \cite{DST}, which is even of interest in
its most simple form of only two sites. Alternatively, this equation
can be derived starting from a many-particle description of a
BEC on a discrete lattice by a Bose-Hubbard Hamiltonian, which leads again
to the DST equation in the mean-field limit. In such a description, however,
the decay has been neglected. In can be re-introduced again by opening the
system. This can be done in various ways, purely phenomenologically by introducing
complex site energies describing decay \cite{08PT,08nhbh_s} or more sophisticatedly
by taking explicitly
the coupling to an environment into account. Recent studies comparing the 
full many particle dynamics with the mean-field approximation consider the
two site system (the open `dimer') phenomenologically  \cite{08PT,08nhbh_s}
or using a master equation for the coupling to the environment \cite{Angl97,Angl01,Trim08}
(see also \cite{Morr08a,Morr08c,Morr08bs} for a study of the related  Lipkin-Meshkov-Glick model).
It should be noted that even the resulting nonhermitian nonlinear two-level
system shows an intricate crossing scenario as discussed, e.g.~, in \cite{06nlnh}.
Two-level systems are often used to model double-well potentials realized in various
BEC experiments \cite{Albi05,Schu05b,Gati06,Gati06b,Foel07}.

The present paper is devoted to a detailed analysis of a simple model system, the
decay behavior of a BEC in a double delta-shell potential.
This open double-well system is on the
one hand simple enough to allow an analytic treatment and closed form approximations
and on the other hand it is flexible enough to investigate characteristic phenomena
observed in nonlinear double-well dynamics, as self-trapping and
the appearance of  new eigenvalues through a saddle node bifurcation \cite{Theo06}
and their modification due to the decay.

The paper is organized as follows: In section~\ref{sec-single} we first discuss the
nonlinear single delta-shell potential and derive simple analytic approximations
for the resonance position and decay rate which are compared with exact numerical
results. These techniques are in section~\ref{sec-double} extended to the double
delta-shell potential where the bifurcation scenario of nonlinear resonance states
is analyzed as well as the decay dynamics. A discrete basis set expansion is used in section~\ref{subsec-DD_Galerkin}
to reduce the system to a finite nonlinear, nonhermitian matrix problem, which yields reasonable
results even for a two state approximation. A Bogoliubov-de-Gennes analysis in section \ref{subsec-DD_BdG} provides
information about the stability of the resonance states.
Additional material concerning computational details is present in an appendix.

\section{Single delta-shell}
\label{sec-single}
At first we consider the case of an open single well, namely the so-called
delta-shell potential
\be
 V(x)= \left\{
                    \begin{array}{cl}
                     +\infty   &    x \le 0 \\
                     (\hbar^2/m) \lambda \, \delta(x-a)  &       x>0\\
                     \end{array}
              \right.\label{single-shell-pot}
\ee
with $\lambda>0$, $a>0$ and repulsive interaction $g>0$. This potential and its generalization to three dimensions have been investigated
in detail in the context of the linear Schr\"odinger equation \cite{Gott66,Dorm80,Kok82,84res1}.
In the context of the GPE it has been considered in \cite{04nls_delta} in which resonance positions
and widths are extracted from real valued wave functions $\psi(x)$
and a resonance is characterized by a maximum of the probability density
$|\psi(x)|^2$ inside the potential region $0 \le x\le a$.

Here we consider nonlinear resonance states, i.e.~eigenstates of the time-independent NLSE
with complex eigenvalues $\mu- \ri \Gamma/2$
\be
    \frac{\hbar^2}{2m} \psi''+(\mu- \ri \Gamma/2 -V)\psi-g|\psi|^2\psi=0
    \label{DS1}
\ee
and purely outgoing (Siegert) boundary conditions (for details see \cite{08nlLorentz}) which enables us to
determine both position $\mu$
and decay width $\Gamma/2$ of the resonances and to construct an approximation that allows an analytical treatment.
Note that the states $\psi(x,t)=\exp\left[-\ri(\mu-\ri \Gamma/2)t/\hbar \right]\,\psi(x)$ do {\it not\/} satisfy the time-dependent
NLSE since their norm is not constant. Instead they provide an adiabatic approximation
to the actual time-dependence (see \cite{Schl06a,Schl06b} and section~\ref{subsec-DD_dynamics}).

The NLSE (\ref{DS1}) with Siegert boundary conditions is solved by the ansatz
\be
 \psi(x)= \left\{
                    \begin{array}{cl}
                      I \sn(\varrho x|p)   & 0 \le x \le a \\
                     C \, \re^{{\rm i}kx}   &     x>a
                    \end{array}
              \right.
\ee
with a Jacobi elliptic $\sn$-function inside and an outgoing plane wave outside the potential well.
The parameters have to satisfy $0 \le p\le 1$,
\be
   \mu -{\rm i}\Gamma/2=\frac{\hbar^2}{2m}\varrho^2(1+p) = \frac{\hbar^2 k^2}{2m}+ g |C|^2
   \label{mu_sn0}
\ee
and
\be
   |I|^2=\frac{\hbar^2\varrho^2p}{gm} \ .
   \label{Iq_sn0}
\ee
We fix the norm of the wavefunction to unity inside the potential region $0 \le x \le a$: 
\be
  1=\int_0^a |\psi(x)|^2 \rd x =\frac{\hbar^2 \varrho}{g m}\left(\varrho a - E(\varrho a|p) \right) \, ,
  \label{DShell_norm}
\ee
where $E(u|p)$ is the incomplete elliptic integral of the second kind.
The matching conditions at $x=a$, $\psi(a+)= \psi(a-)=\psi(a)$, $\psi'(a+)- \psi'(a-)=2\lambda \psi(a)$, yield
\begin{eqnarray}
   I \sn(\varrho a|p)= C\, \re^{\ri k a} \\
   I \varrho \, \cn(\varrho a|p) \dn(\varrho a|p) + 2 \lambda I \sn(\varrho a|p) 
   = \ri k C\, \re^{\ri k a} \, . \label{DS2}
\end{eqnarray}
For a strongly repulsive delta-function at $x=a$ we have $|k| \ll \lambda$ and 
$\sn(\varrho a|p) \approx 0$ so that the decay of the lowest resonances is weak.
Thus we neglect the imaginary part $-\Gamma/2$ of the eigenvalue in equation ~(\ref{mu_sn0}) and (\ref{DS2}).
For $\lambda \rightarrow \infty$ the wave number $\varrho$ is given by 
$\varrho=2 n K(p)/a$ ($ K(p)$ is the complete elliptic integral of the first kind) 
so that for weak decay we can assume
\be
   \varrho=\frac{2 n K(p)}{a}+\delta
   \label{rho_plus_delta}
\ee
with $\delta <0$, $|\delta \cdot a| \ll 1$. Expanding the real part of equation (\ref{DS2}) up to second 
order in $\delta$ we obtain
\be
   \delta =\frac{2 \lambda a +1}{2 n K(p)(1+p)a}-\sqrt{\left(\frac{2 \lambda a +1}{2 n K(p)(1+p)a}\right)^2
   +\frac{2}{(1+p)a^2}} \, .
   \label{delta_DShell}
\ee
For a given value of the Jacobi elliptic parameter $0 \le p \le 1$
the real part of the eigenvalue is determined by the equations (\ref{mu_sn0}), (\ref{rho_plus_delta}) and 
(\ref{delta_DShell}). From the normalization condition (\ref{DShell_norm}) the interaction strength is given by
\be
 g =\frac{\hbar^2 \varrho}{m}\left(\varrho a - E(\varrho a|p) \right)\,.
\ee
The imaginary part of the eigenvalue can be estimated using the Siegert relation \cite{Sieg39,08nlLorentz}
\be
   \Gamma/2=\frac{\hbar^2 k}{2m} \frac{|\psi(a)|^2}{ \int_0^a |\psi(x)|^2 \rd x}
   =\frac{\hbar^2 k \varrho p}{2m} \frac{\sn^2(\varrho a|p)}{\varrho a - E(\varrho a|p)} \, .
\ee

\begin{figure}[htb]
\centering
\includegraphics[width=7cm,  angle=0]{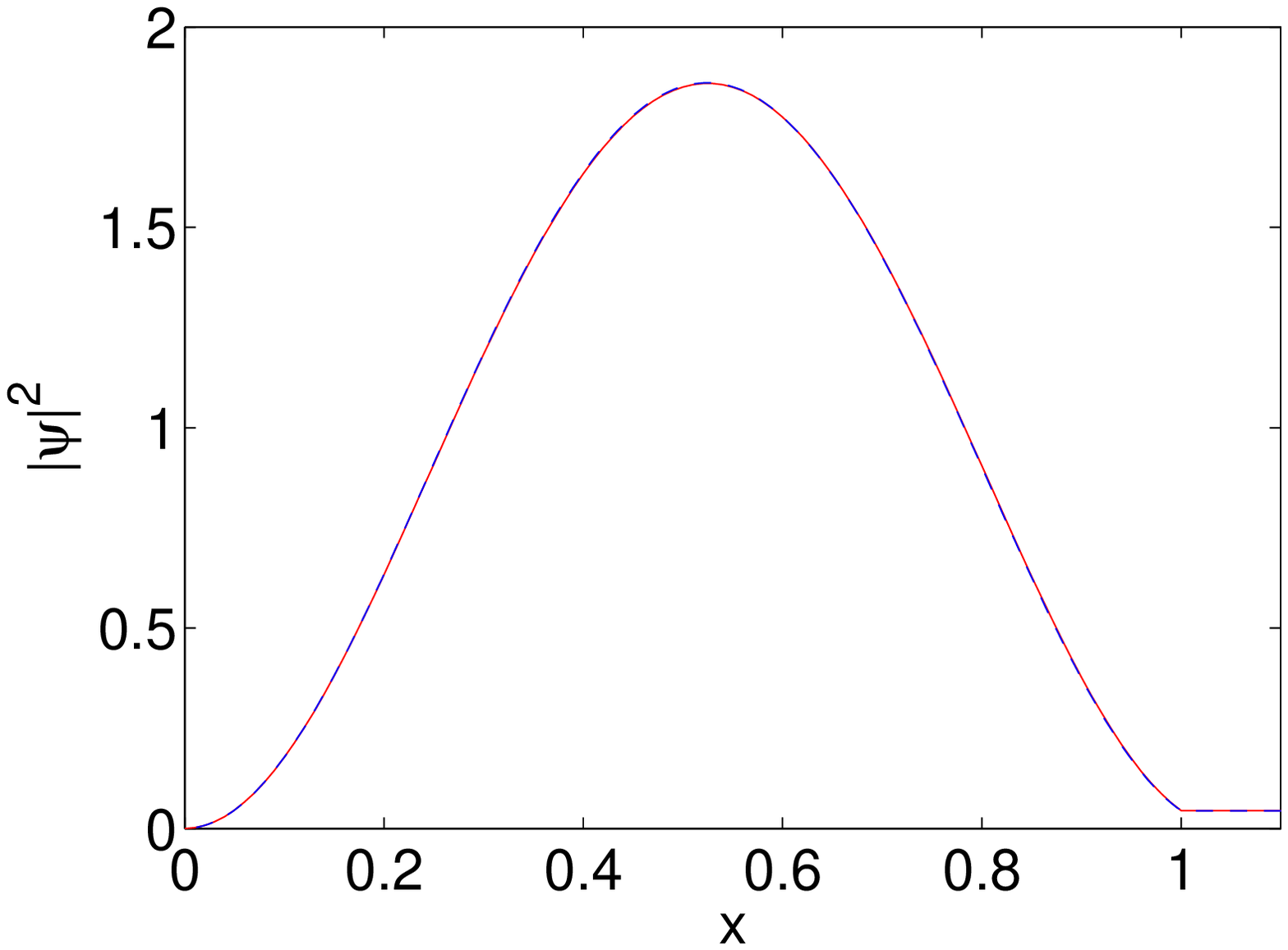}
\includegraphics[width=7cm,  angle=0]{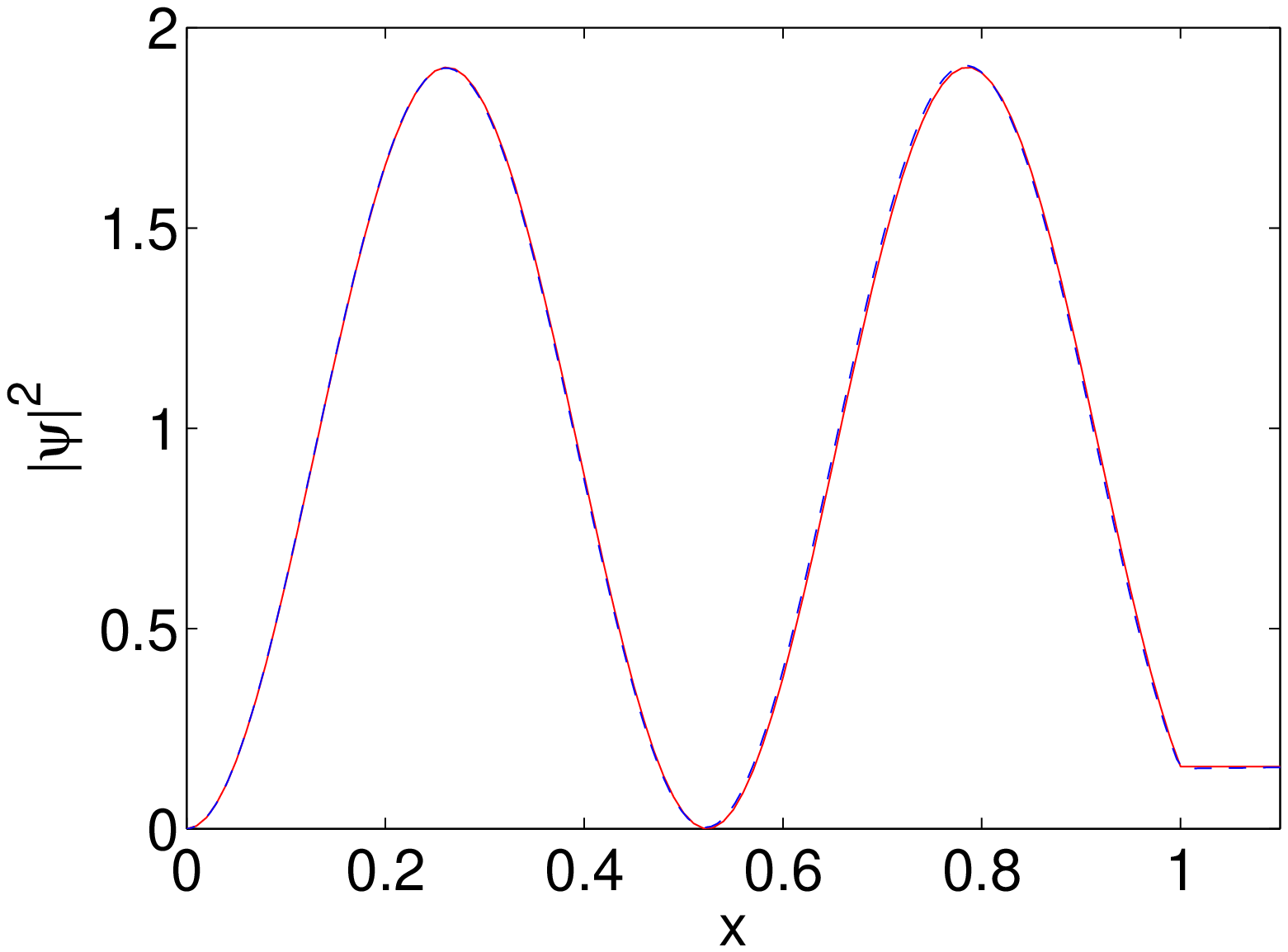}
\caption{\label{fig-DShell} Resonance wave function for the ground and first excited state of
a delta-shell potential (parameters $\lambda=10$, $a=1$ and nonlinearity $g=1$; units $\hbar=m=1$). 
Left: ground state, $n=1$, right: first excited state, $n=2$.
The exact numerical solutions obtained by the CAP method (dashed blue) 
are in excellent agreement with the analytical approximations (solid red).}
\end{figure}

Exact results can be obtained from a numerical calculation combining a complex
absorbing potential (CAP) with a grid relaxation. In these computations we use the approximate resonances derived above as starting values. 
Complex absorbing potentials were shown to be equivalent to exterior complex scaling \cite{Mois98} and successfully
applied to resonances of the GPE \cite{Mois04a}.
A detailed discussion of our particular implementation as well as its application to nonlinear Wannier-Stark resonances
will be presented elsewhere \cite{08nlws}.

As an example, figure~\ref{fig-DShell}
shows the wavefunctions for 
the ground state and the first excited state of the delta-shell potential (\ref{single-shell-pot})
 with $\lambda=10$, $a=1$  and a nonlinearity of $g=1$ (units with $\hbar=1=m$ are used throughout this
 paper) and
table~\ref{tab-DShell} lists the resonance energies. In addition the analytical approximations
derived above are given.  Good agreement is observed.


\begin{table}[htbp]
\centering
\begin{tabular}{c|cc|cc}
 $n$ & $\mu_{\rm A}$ & $\mu_{\rm CAP}$  &
  $\Gamma_{\rm A}/2$ & $\Gamma_{\rm CAP}/2$\\
  \hline
  1 & 5.8950 & 5.9043 & 0.0759 & 0.0751  \\
  2 & 19.4138 & 19.5065 & 0.4810 &0.4680  \\
\end{tabular}
\caption{ Resonance energies $\mu_n-\ri \Gamma_n/2$ for the two lowest resonances of the 
delta-shell potential (parameters $\lambda=10$, $a=1$ and nonlinearity $g=1$) calculated
analytically (approximation A) and numerically using a grid relaxation method with complex absorbing 
potentials (CAP).}
\label{tab-DShell}
\end{table}

\section{Double delta-shell}
\label{sec-double}
An open double well allows to study the influence of the decay on 
characteristic nonlinear phenomena like self-trapping. Here we investigate a simple
case, namely the double-delta-shell potential
\be
 V(x)= \left\{
                    \begin{array}{cl}
                     +\infty   &    x \le 0 \\
                     (\hbar^2/m) \left[\lambda_b\, \delta(x-b) + \lambda_a \, \delta(x-a) \right]&   x>0\\
                     \end{array}
              \right. \label{DDShell_pot}
\ee
which consists of an infinitely high wall and two repulsive delta barriers
with strength $\lambda_a>0$, $\lambda_b>0$ located at  $a>b>0$. This potential
possesses two simpler limits: For $\lambda_a \rightarrow 0$ or $\lambda_b \rightarrow 0$ 
we recover the single delta-shell potential discussed in section~\ref{sec-single}. In the
limit $\lambda_a \rightarrow \infty$ the system is closed yielding a coherent nonlinear tunneling 
oscillation between two (asymmetric) potential wells \cite{Theo06,Khom07}.
For finite values of $\lambda_a$ we therefore observe nonhermitian generalizations of these simple
cases.
The considerations in the following subsection
essentially follow those from the preceding section~for the single delta-shell potential.
We first consider stationary resonance states followed by a discussion of the dynamics and a
stability analysis. As in the preceding section we concentrate on repulsive nonlinearities $g>0$.
\subsection{Stationary States}
\label{subsec-DD_stationary}
The NLSE (\ref{DS1}) with the potential (\ref{DDShell_pot}) and Siegert boundary conditions is solved 
by the ansatz
\be
 \psi(x)= \left\{
                    \begin{array}{cl}
                      I_1 \sn(\varrho_1 x|p_1)   & 0 \le x \le b \\
                      I_2 \sn(\varrho_2 x+\vartheta|p_2)   & b < x \le a \\
                     C \, \re^{{\rm i}kx}                              &     x>a
                    \end{array}
              \right.
\ee
with
\be
   \mu-\ri \Gamma/2 =\frac{\hbar^2}{2m}\varrho_1^2(1+p_1) = \frac{\hbar^2}{2m}\varrho_2^2(1+p_2)=\frac{\hbar^2 k^2}{2m}+ g |C|^2 \, ,
   \label{mu_sn}
\ee
\be
   |I_1|^2=\frac{\hbar^2\varrho_1^2p_1}{gm} \, , \quad |I_2|^2=\frac{\hbar^2\varrho_2^2p_2}{gm} \
   \label{Iq_sn}
\ee
and $0 \le p \le 1$.
At $x=a$ we obtain the matching conditions
\begin{eqnarray}
   I_2 \sn(\varrho_2 a +\vartheta |p_2)= C\, \re^{\i k a} \\
   I_2 \varrho_2 \cn(\varrho_2 a +\vartheta|p_2) \dn(\varrho_2 a+\vartheta|p_2) 
   + 2 \lambda_a I_2 \sn(\varrho_2 a+\vartheta|p_2) = i k C\, \re^{\i k a} \label{DD2} \, .
\end{eqnarray}
Due to the repulsive delta-function at $x=a$ we have $k \ll \lambda_a$ and $\sn(\varrho_2 a 
+\vartheta|p_2) \approx 0$. 
As in the previous section~we neglect the decay coefficient $\Gamma/2$  in equation (\ref{DD2}). Furthermore
we assume 
$\varrho_2 a+\vartheta=4 n K(p_2) + \delta \cdot a$ with $\delta \cdot a <0$, $|\delta \cdot a| \ll 1$. 
Expanding the real part of equation (\ref{DD2}) up to second order in $\delta$ we obtain
\be
   \delta =\frac{2 \lambda_a a}{\varrho_2(1+p_2)a^2}-\sqrt{\left(\frac{2 \lambda_a a}{\varrho_2(1+p_2)a^2}\right)^2
   +\frac{2}{(1+p_2)a^2}} \, . \label{DD-delta}
\ee
so that the phase shift is given by $\vartheta=4 n K(p_2)+ \delta \cdot a - \varrho_2 a$.

The matching conditions at $x=b$ read
\begin{eqnarray}
   I_1 \sn(u_1|p_1)=I_2 \sn(u_2|p_2) \label{DD3} \, ,\\
   I_1 \varrho_1 \cn(u_1|p_1)\dn(u_1|p1)+2 \lambda_b I_1 \sn(u_1|p_1)=I_2 \varrho_2 \cn(u_2|p_2) \dn(u_2|p_2) \label{DD4}
\end{eqnarray}
where $u_1=\varrho_1 b$, $u_2=\varrho_2 b +\vartheta$.
By inserting equation (\ref{DD3}) into (\ref{DD4}) and using the relations between the squares of the Jacobi elliptic
functions we arrive at
\begin{eqnarray}
 \frac{p_1}{(p_1+1)^2}=\frac{p_2}{(p_2+1)^2}&-&\frac{p_2}{(p_2+1)^{3/2}}\frac{4 \lambda_b}{\sqrt{2 \mu}}\,\sn(u_2|p_2)\cn(u_2|p_2)\dn(u_2|p_2) \nonumber \\
 &+& \frac{2 \lambda_b^2}{\mu}\frac{p_2}{p_2+1}\,\sn(u_2|p_2)^2 =:F(p_2,\mu)
\end{eqnarray}
so that the parameter $p_1$ is given by
\be
   p_1=\frac{1}{2F}-1-\sqrt{\left(\frac{1}{2F}-1 \right)^2-1} \label{DD-p} \, .
\ee
From equations (\ref{DD-delta}) -- (\ref{DD-p}) all relevant quantities are known in terms of $p_2$ and $\mu$. 
To calculate the eigenvalues we solve equation (\ref{DD4}) for a given value of $\mu$ and thus obtain up to
four solutions for $p_2$ in $[0,1]$.

From the normalization condition $\int_0^a |\psi(x)|^2 \rd x =1$ we can determine the interaction parameter
\begin{eqnarray}
g=&\frac{\hbar^2}{m}&\Big[\varrho_1\left(\varrho_1 b-E(\varrho_1 b|p_1) \right) \nonumber \\
  &+&\varrho_2 \left(\varrho_2 (a-b)+E(\varrho_2 b+\vartheta|p_2)-E(\varrho_2 a+\vartheta|p_2) \right)\Big] \, .
\end{eqnarray}
The decay coefficient is again estimated by the Siegert relation
\be
   \Gamma/2=\frac{\hbar^2 k}{2m} \frac{|\psi(a)|^2}{ \int_0^a |\psi(x)|^2 \rd x} \, . \label{DD-Siegert}
\ee

\begin{figure}[htb]
\centering
\includegraphics[width=7cm,  angle=0]{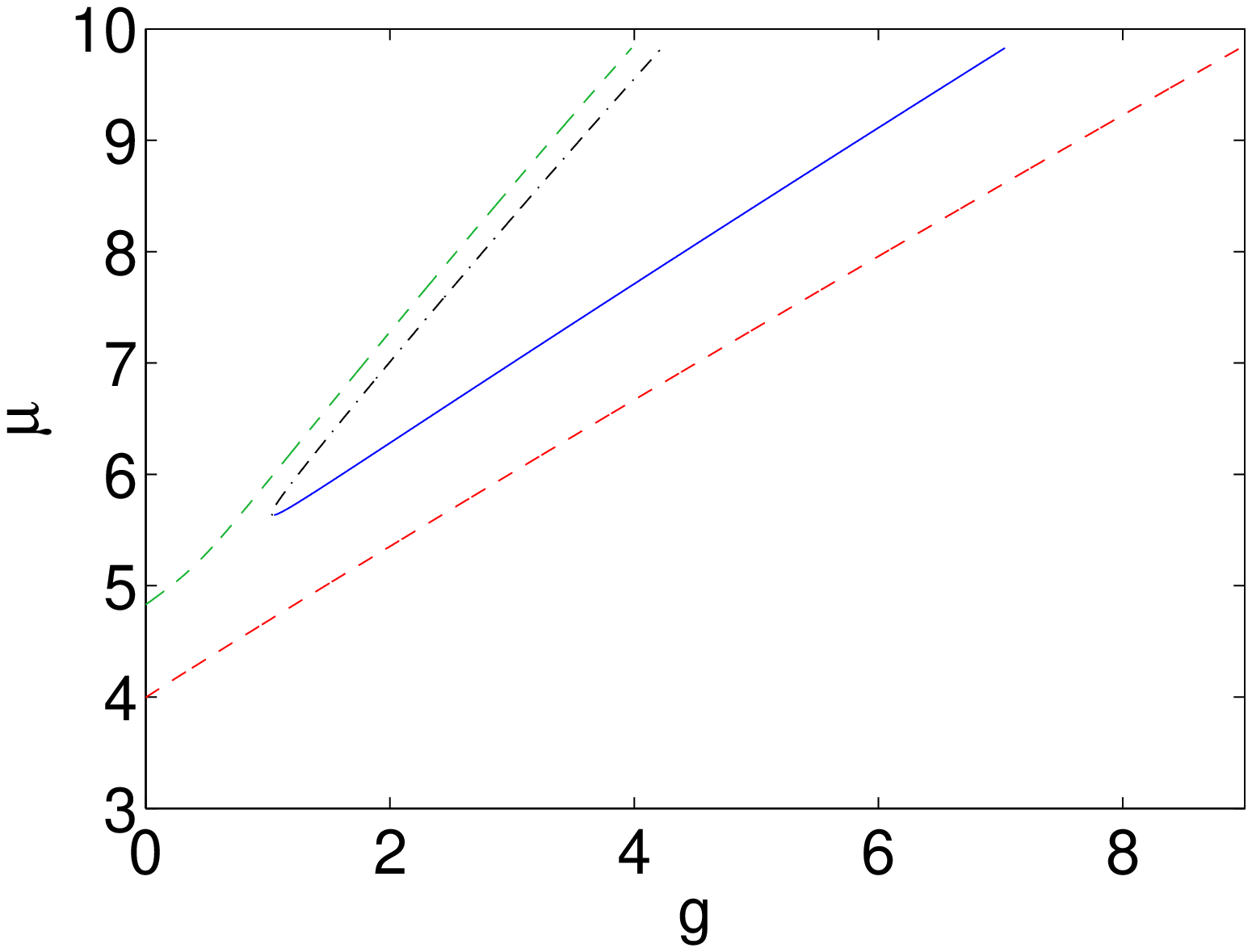}
\includegraphics[width=7cm,  angle=0]{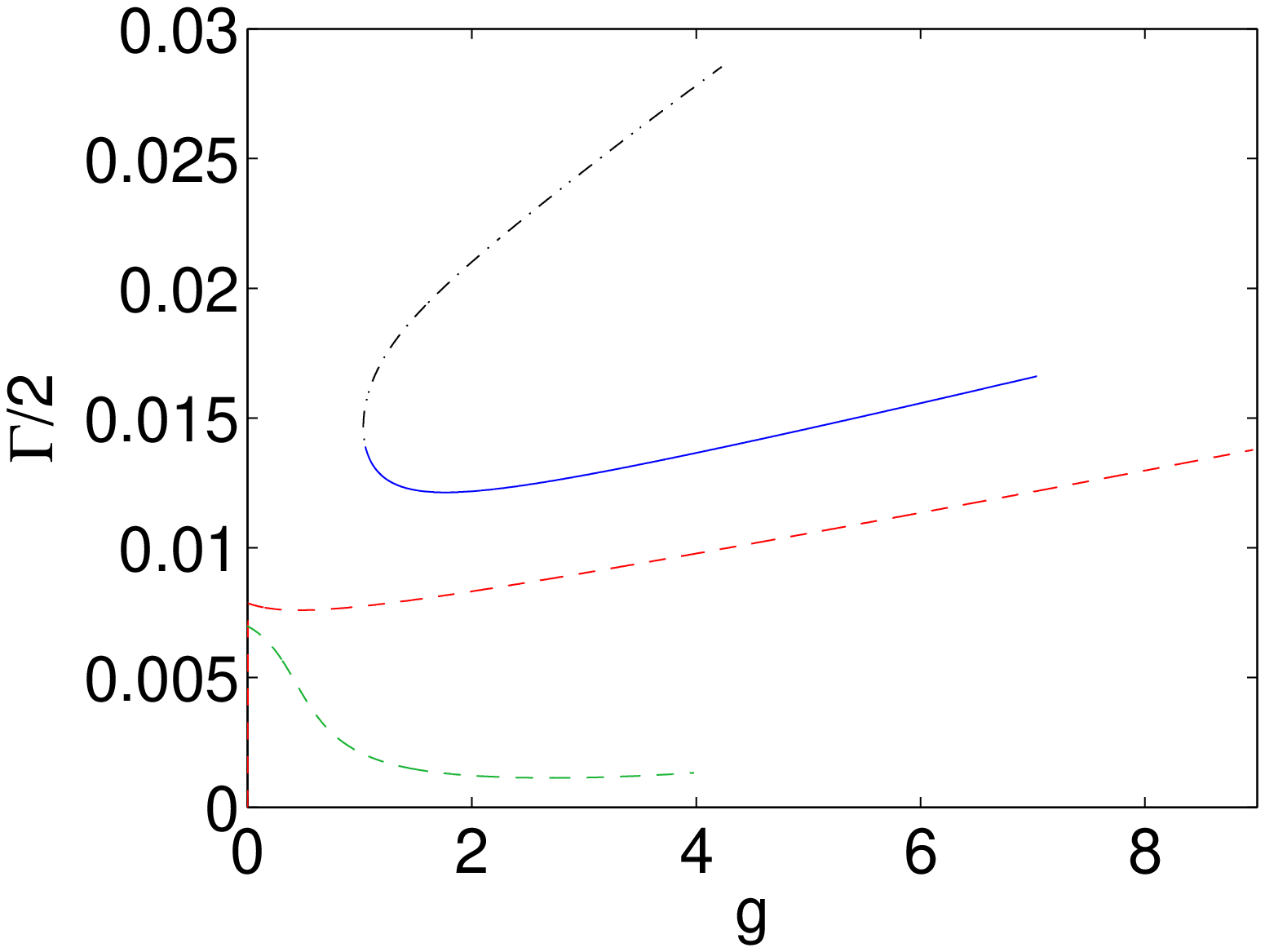}\\
\caption{\label{fig-DD1} {Analytically calculated bifurcation diagrams for the parameters $\lambda_b=10$,
$\lambda_a=20$, $b=1$ and $a=2$. Left panel: Chemical potentials. Right panel: Decay rates.}}
\end{figure}

\begin{figure}[htb]
\centering
\includegraphics[width=7cm,  angle=0]{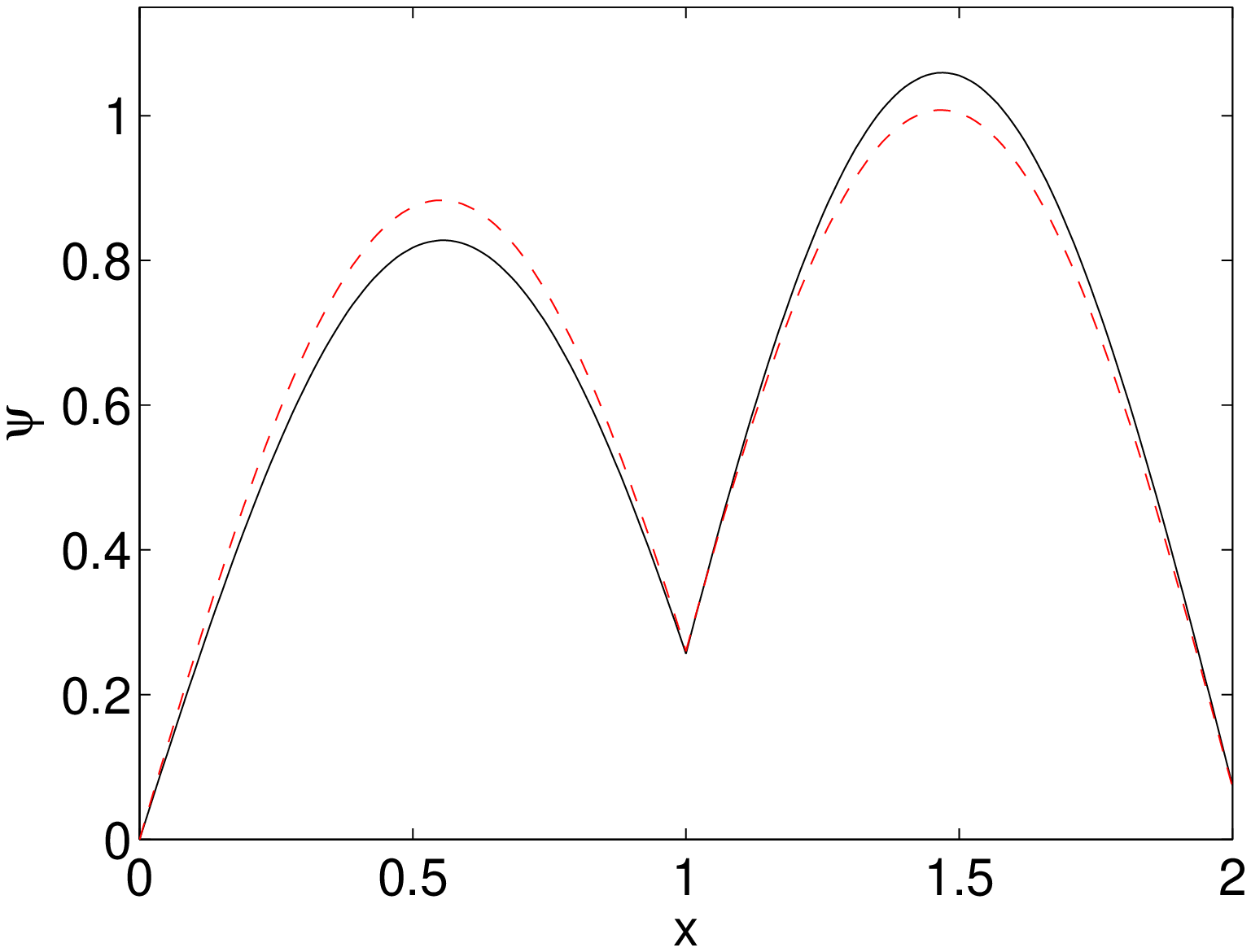}
\includegraphics[width=7cm,  angle=0]{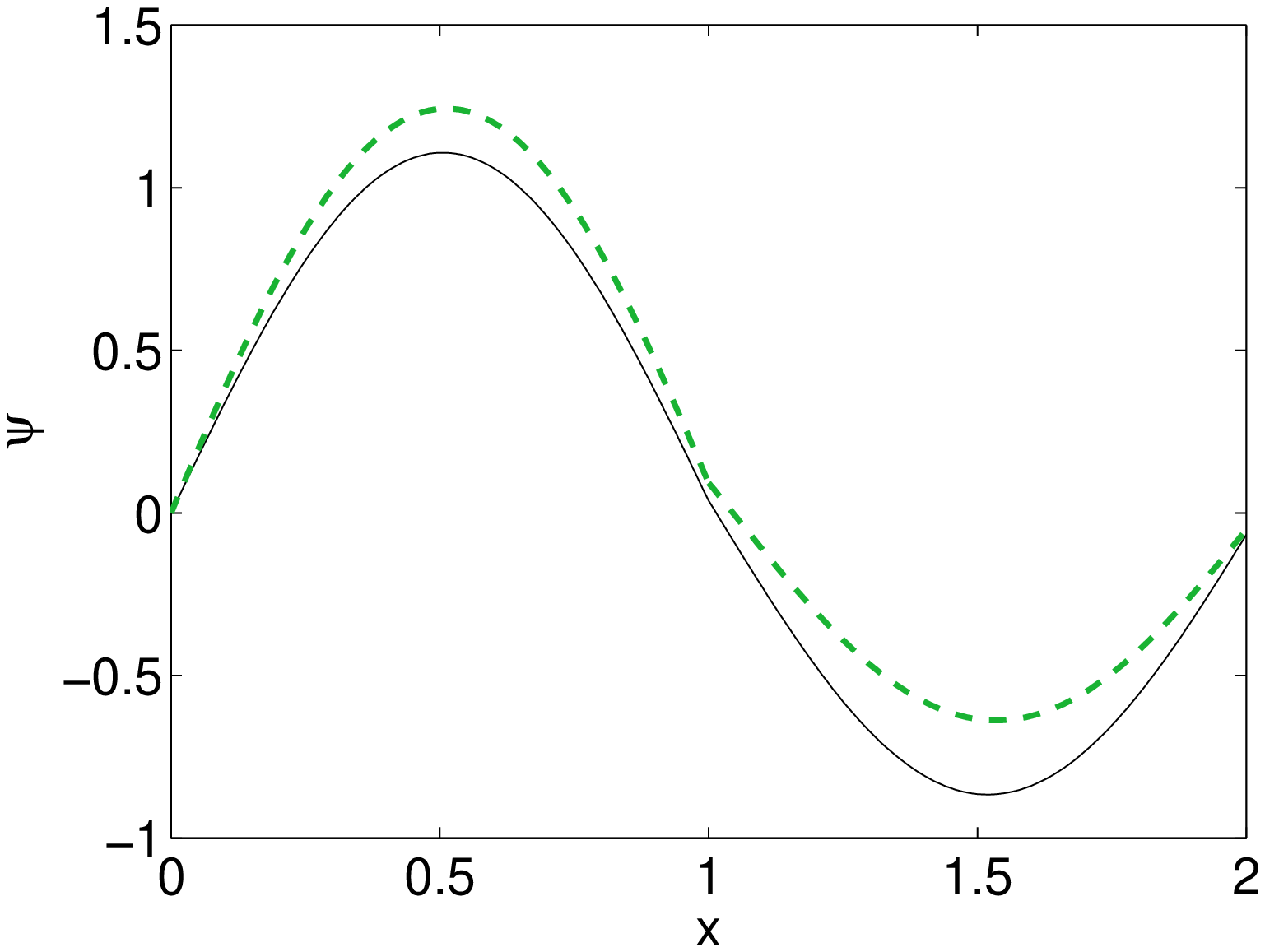}
\caption{\label{fig-DD_g0_5_g0} {Wavefunctions of the two lowest states for $\lambda_b=10$, $\lambda_a=20$, 
$b=1$, $a=2$. Left panel: ground state for $g=0$ (solid black) and $g=0.5$ (dashed red).
Right panel: first excited state for $g=0$ (solid black) and $g=0.5$ (dashed green).}}
\end{figure}

\begin{figure}[htb]
\centering
\includegraphics[width=7cm,  angle=0]{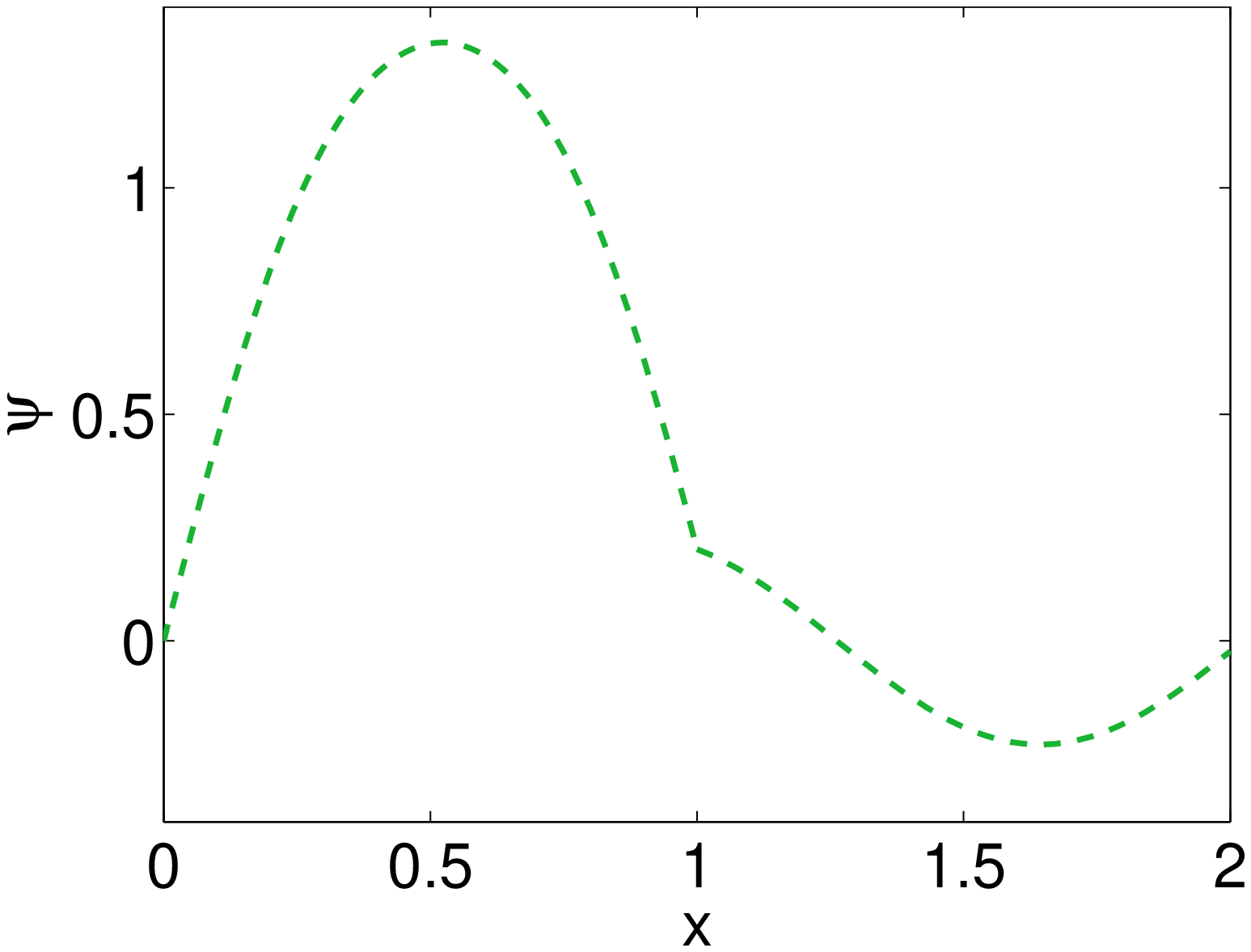}
\includegraphics[width=7cm,  angle=0]{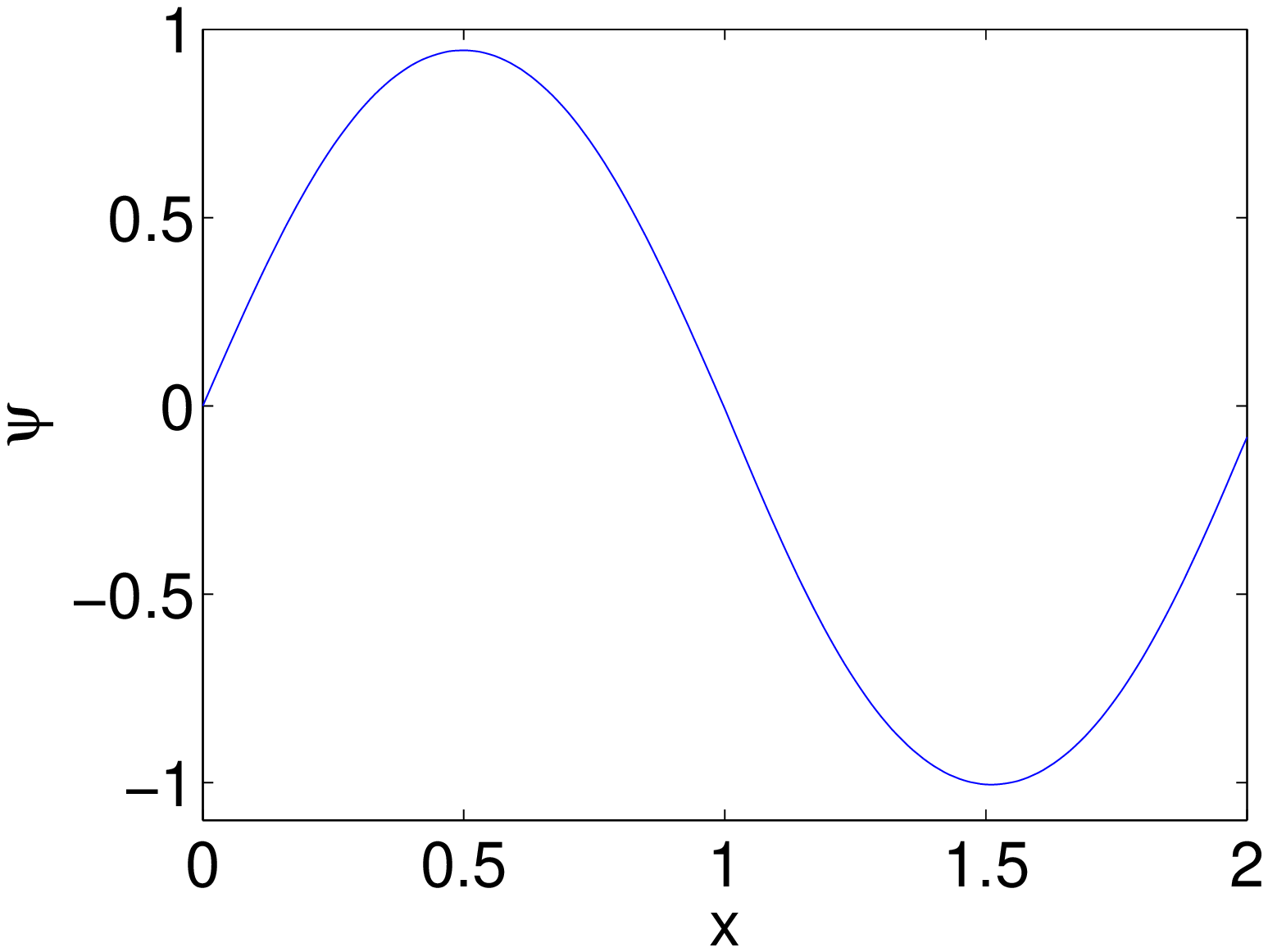}\\
\includegraphics[width=7cm,  angle=0]{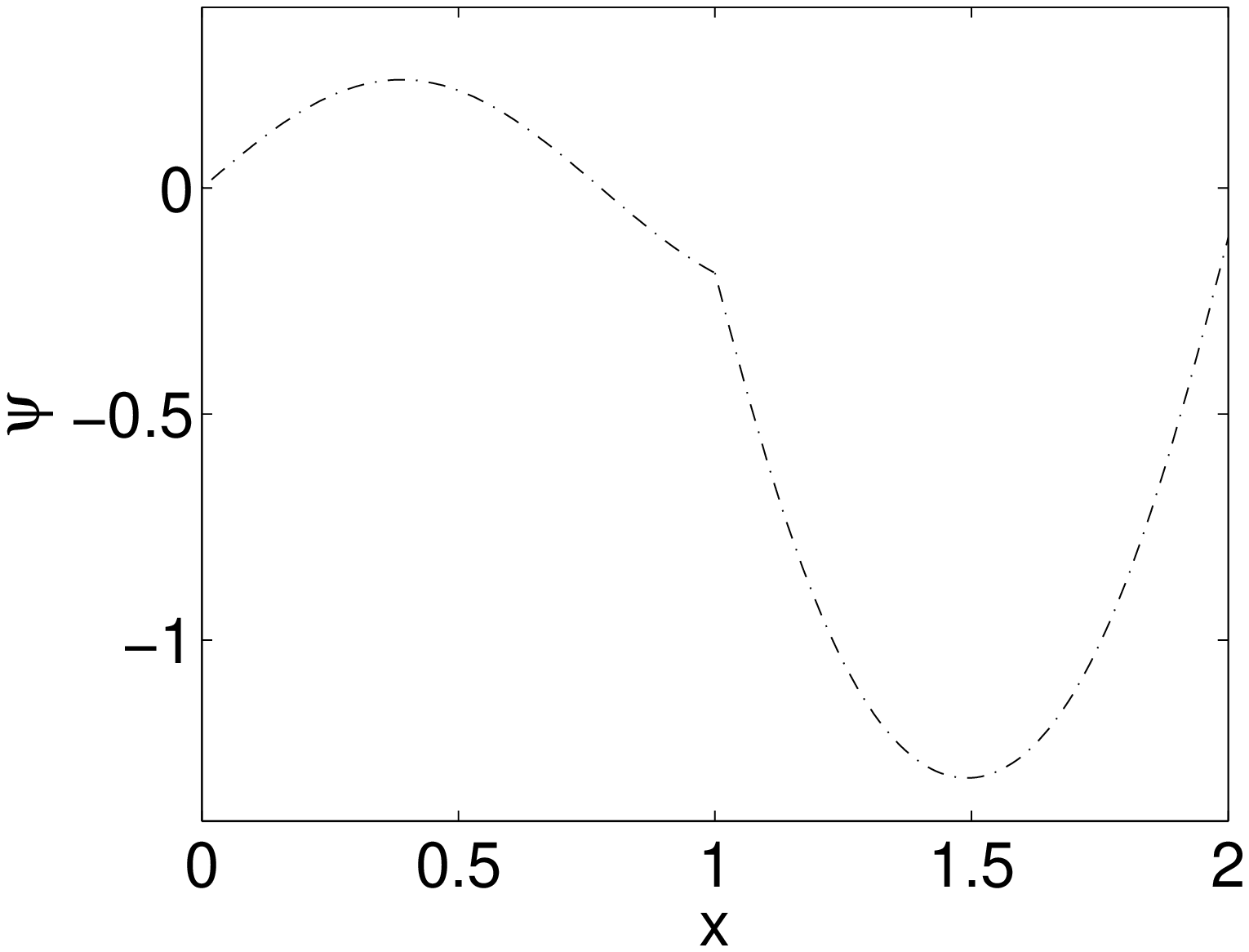}
\includegraphics[width=7cm,  angle=0]{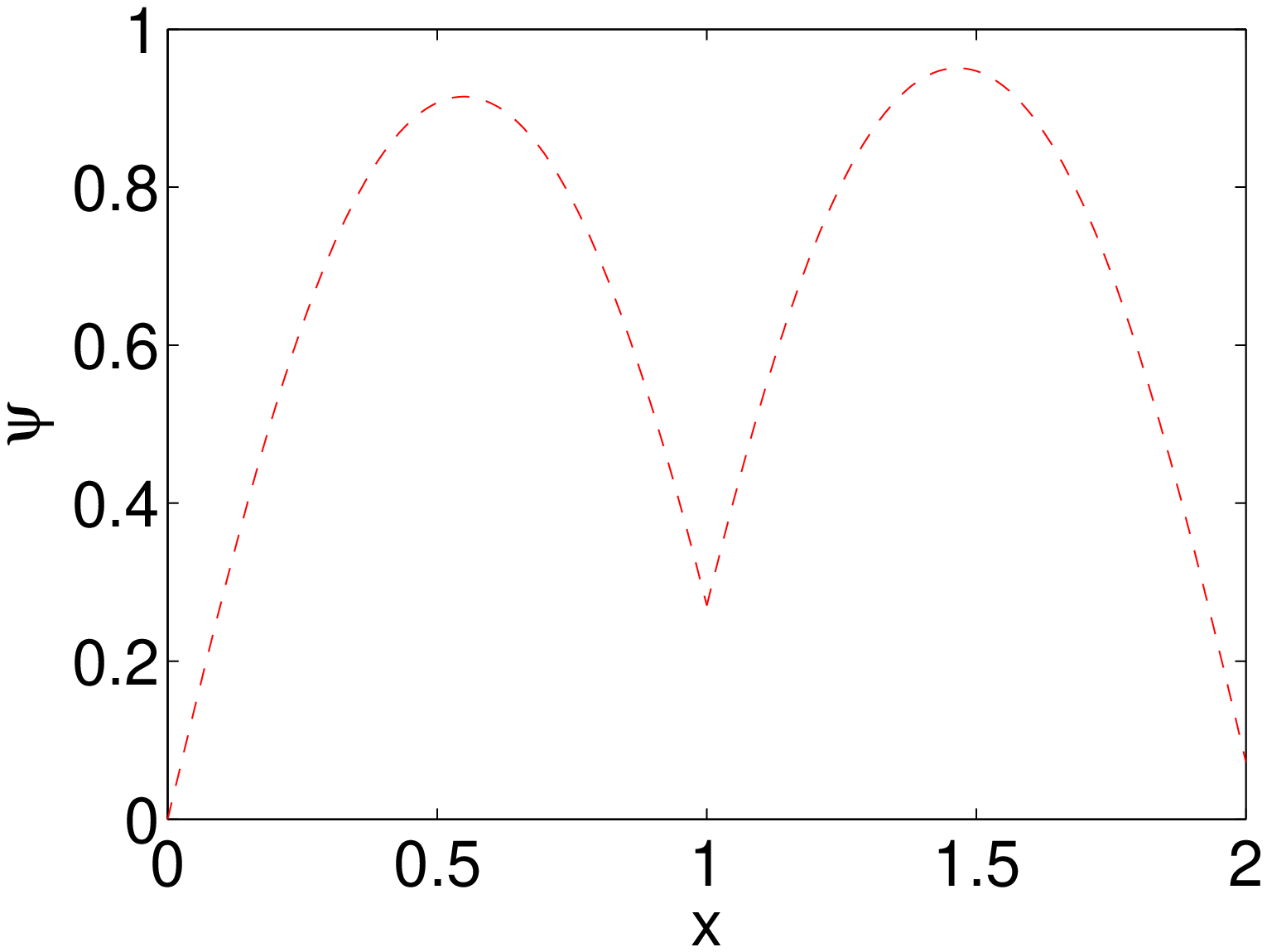}
\caption{\label{fig-DD2} {Analytically calculated wavefunctions for $\lambda_b=10$, $\lambda_a=20$,
$b=1$, $a=2$ and $g=3$. Upper left: {autochtonous} self-trapping state (AuT), lower left:
{allochtonous} self-trapping state (AlT), upper right: allochtonous almost antisymmetric state (Al-),
lower right: autochtonous almost symmetric state (Au+).}} 
\end{figure}

\begin{figure}[htb]
\centering
\includegraphics[width=7cm,  angle=0]{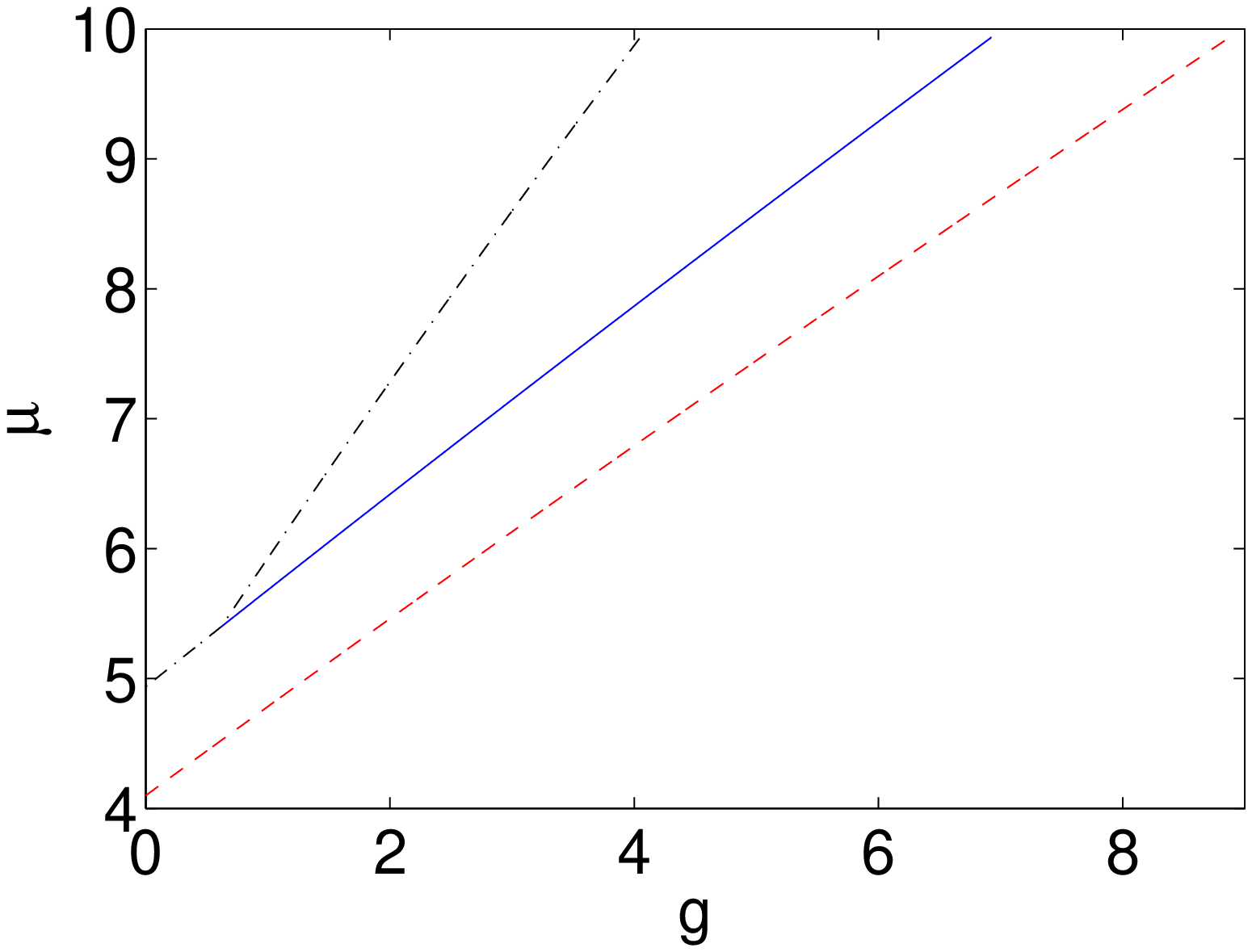}
\includegraphics[width=7cm,  angle=0]{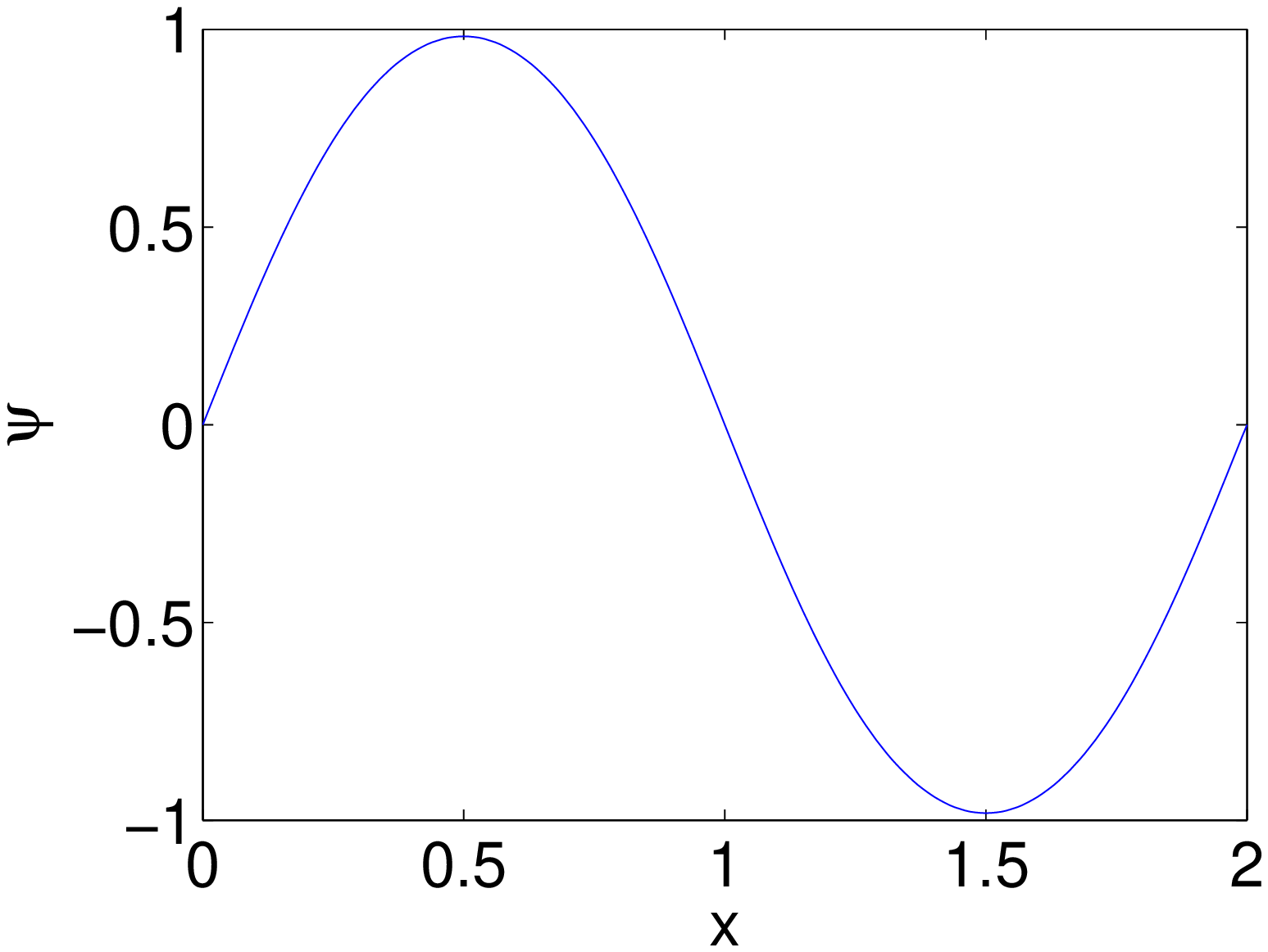}\\
\includegraphics[width=7cm,  angle=0]{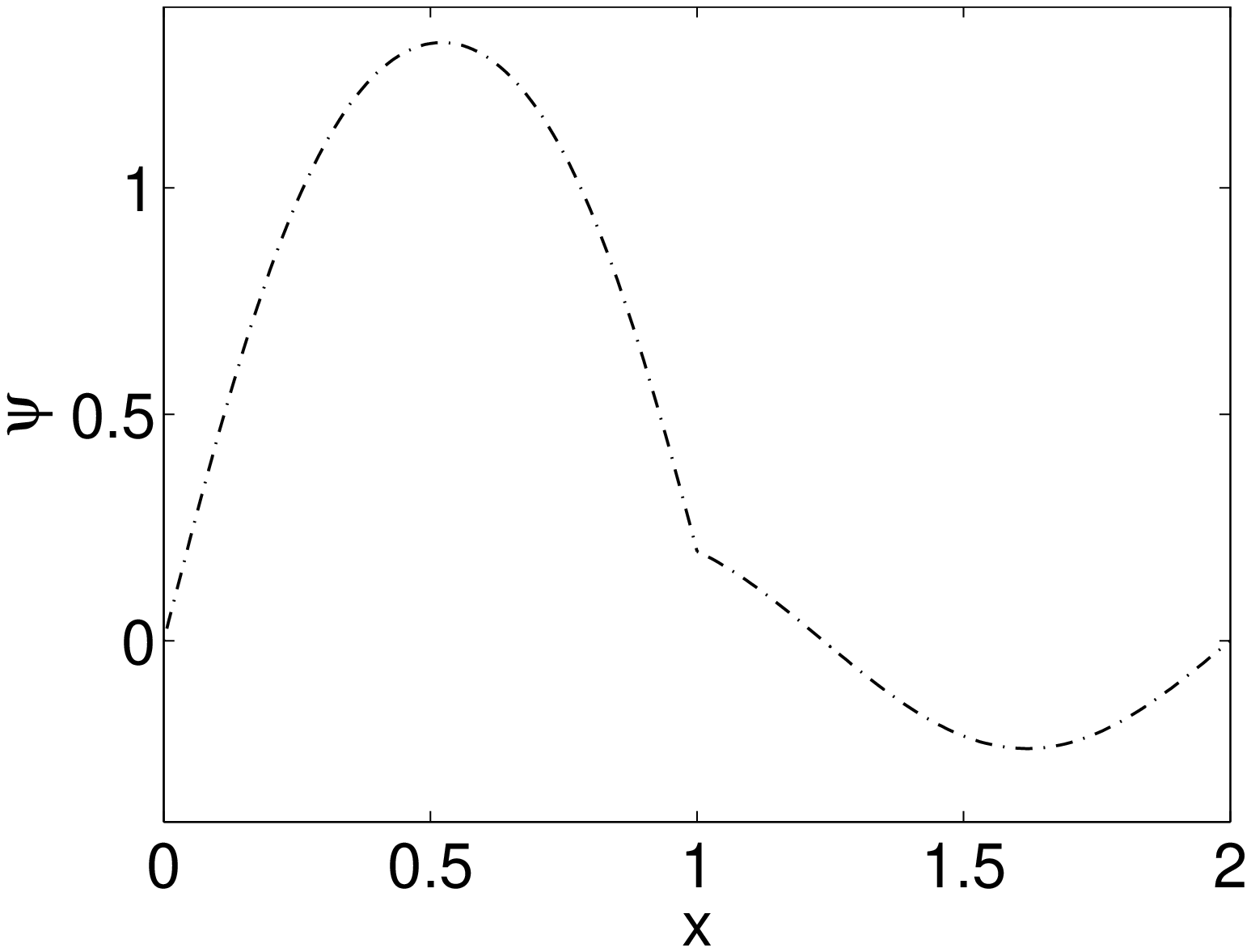}
\includegraphics[width=7cm,  angle=0]{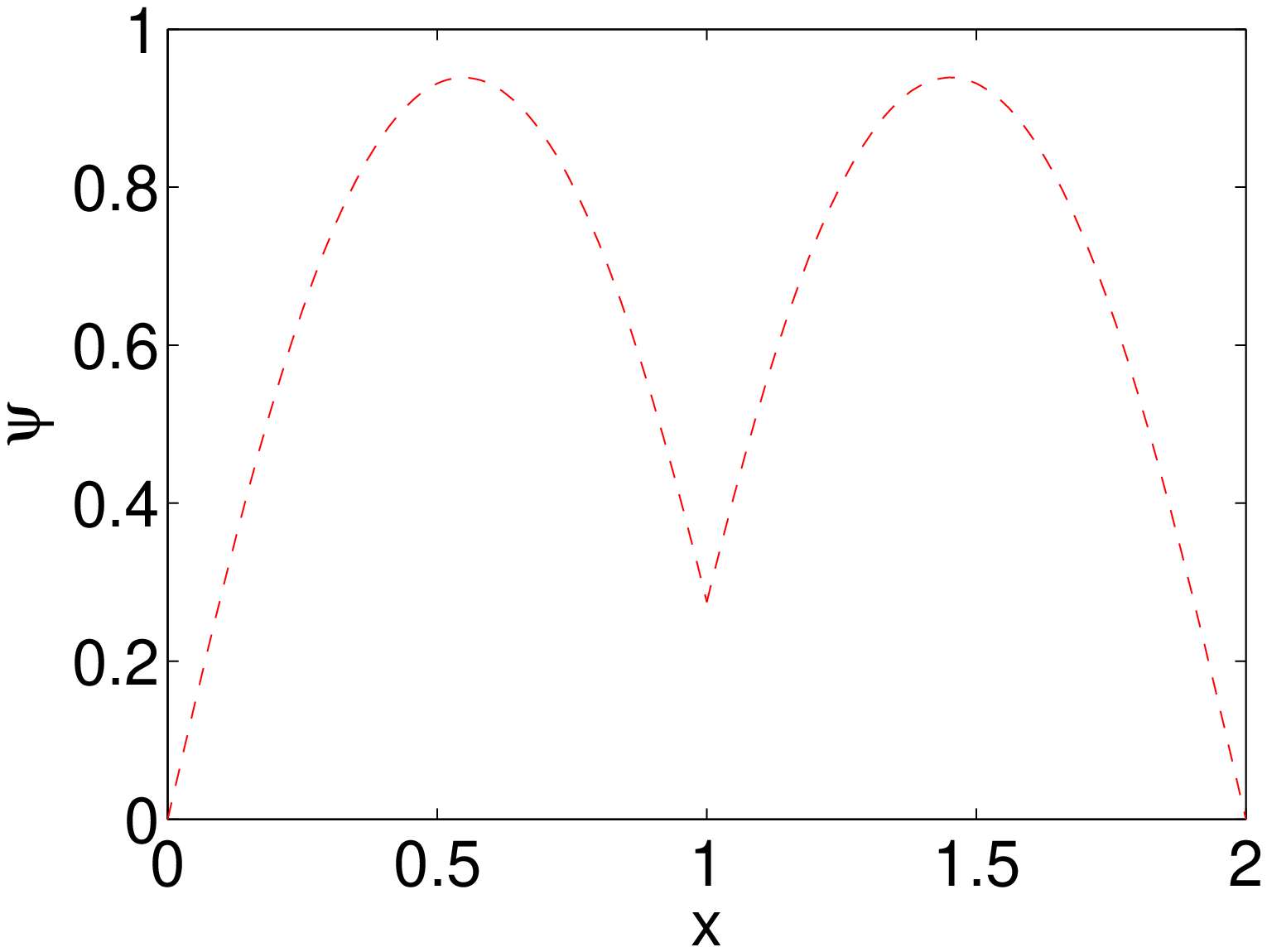}
\caption{\label{fig-DD_u} {Analytical results for $\lambda_b=10$, $\lambda_a \rightarrow \infty$, $b=1$,
$a=2$. Upper left panel: Bifurcation diagram. The other panels show eigenstates for $g=3$.  Lower left: self-trapping state. Upper right:antisymmetric state. Lower right: symmetric state.}}
\end{figure}
\begin{figure}[htb]
\centering
\includegraphics[width=7cm,  angle=0]{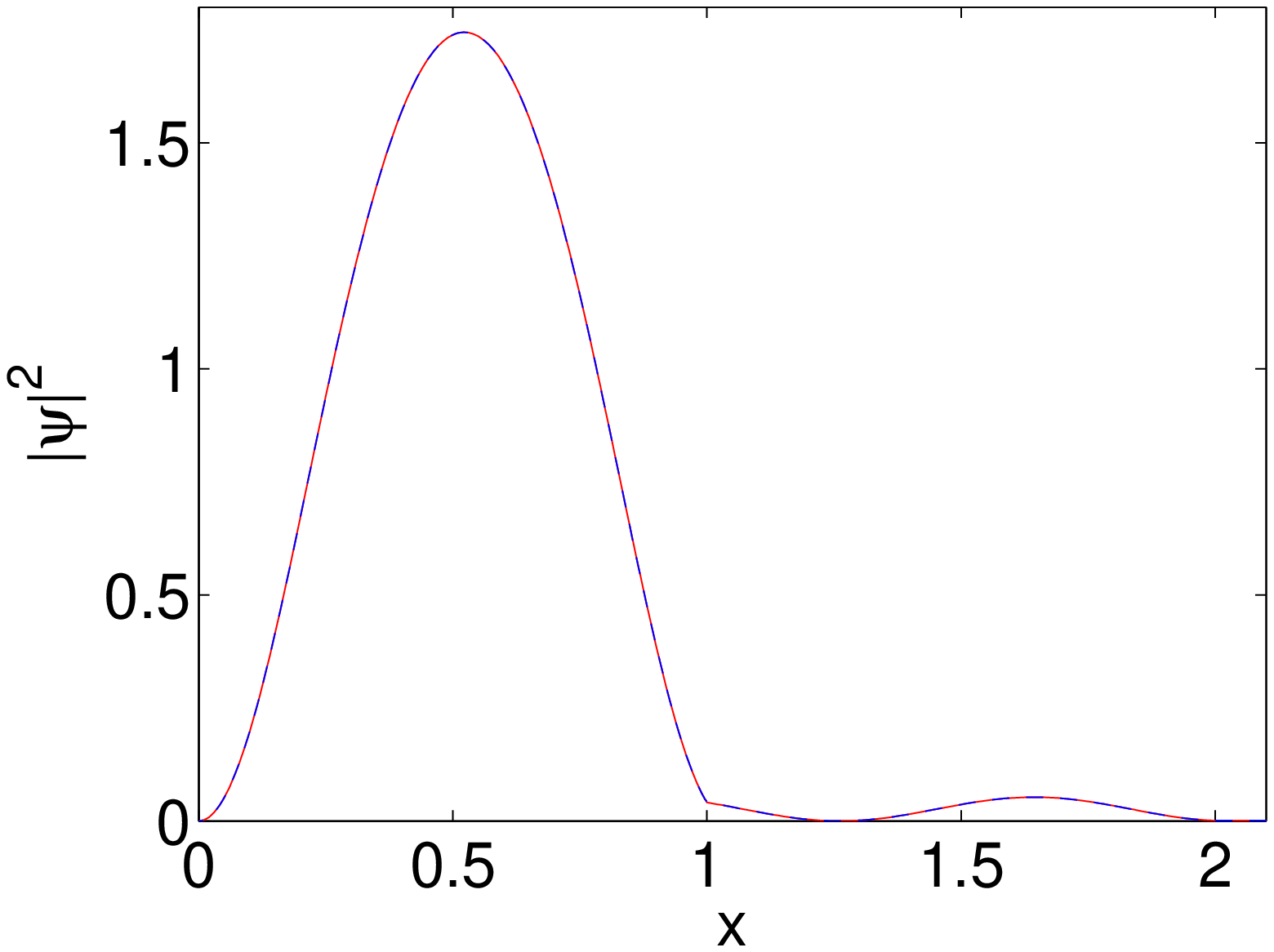}
\includegraphics[width=7cm,  angle=0]{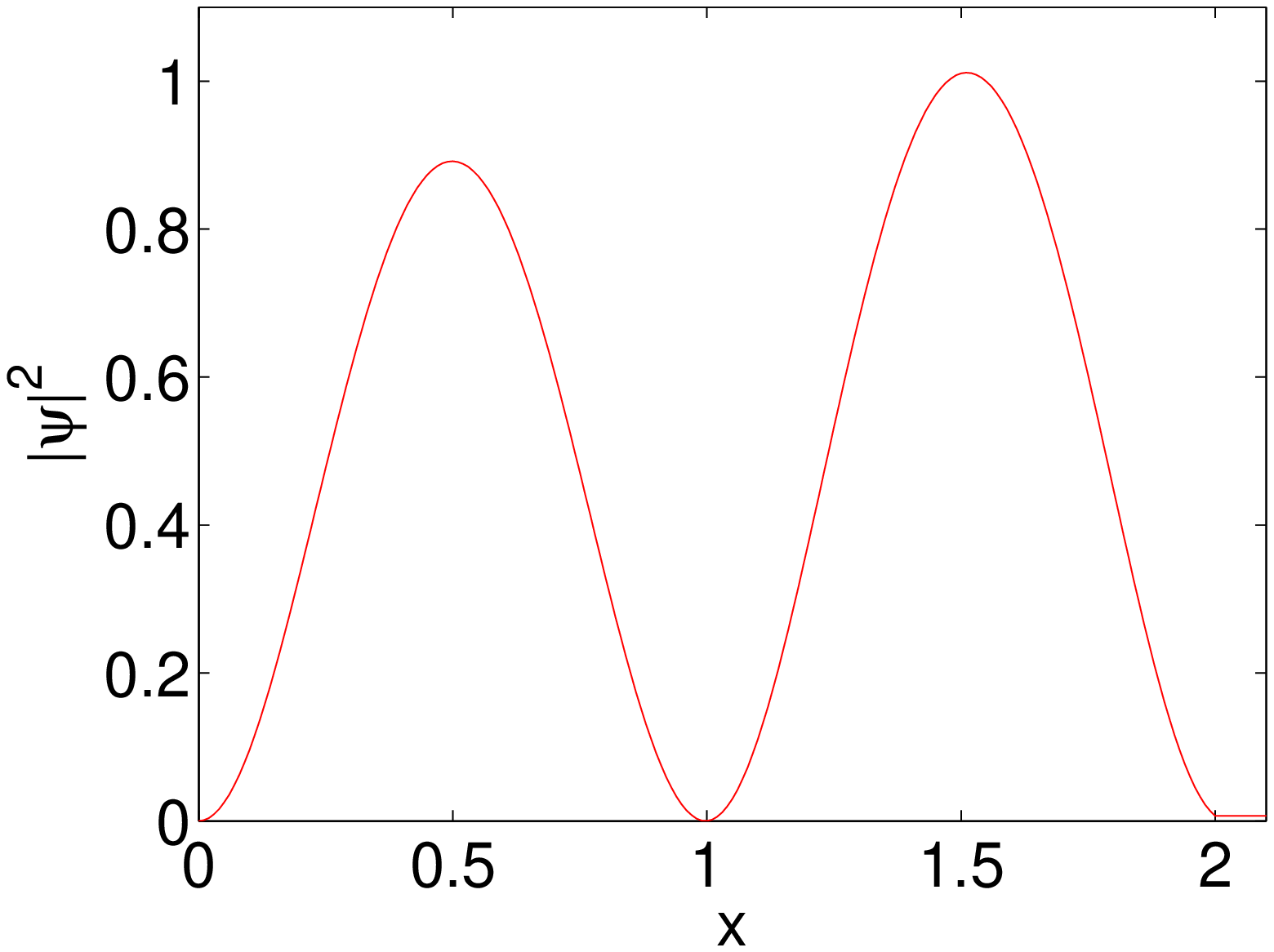}\\
\includegraphics[width=7cm,  angle=0]{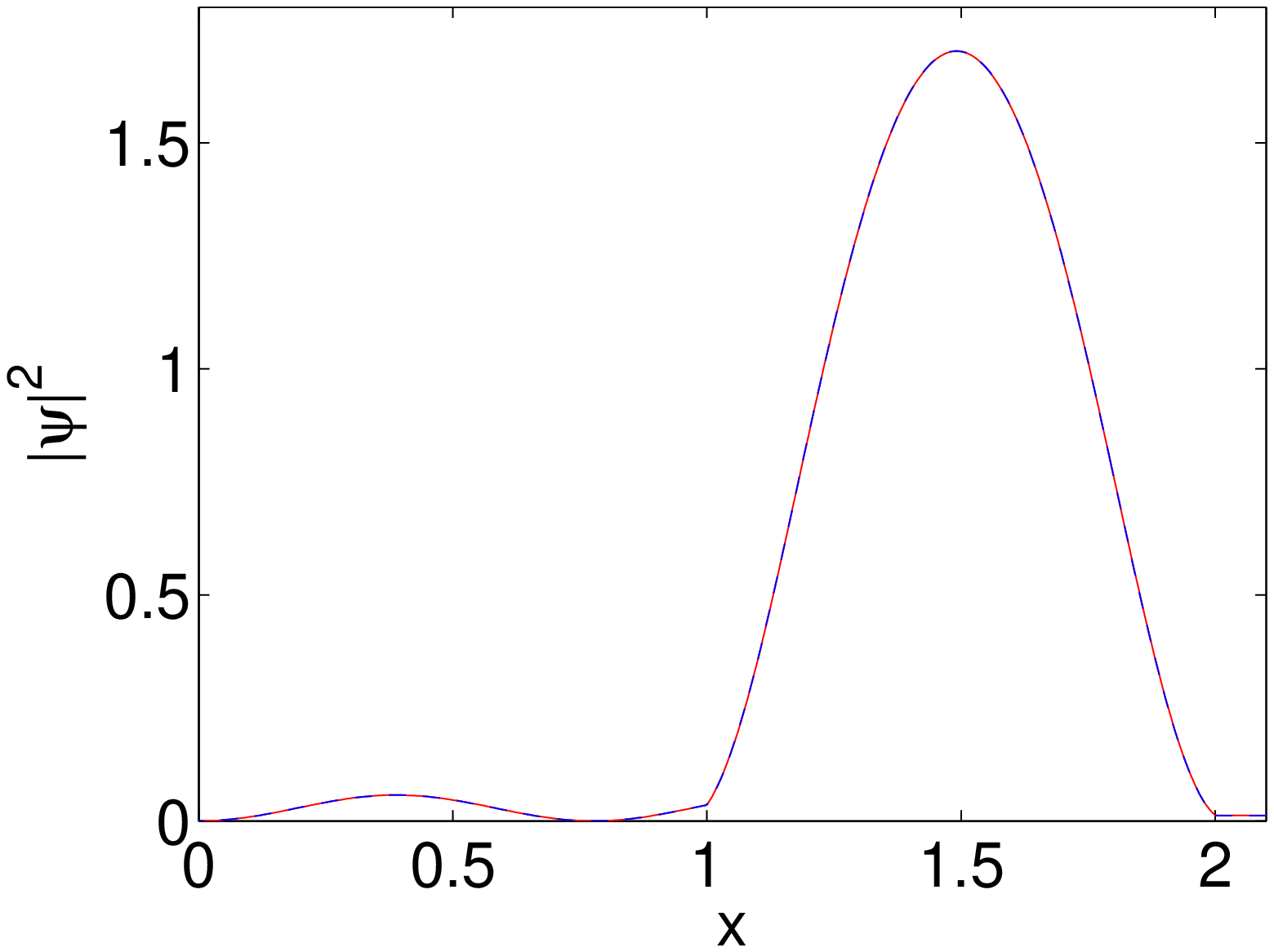}
\includegraphics[width=7cm,  angle=0]{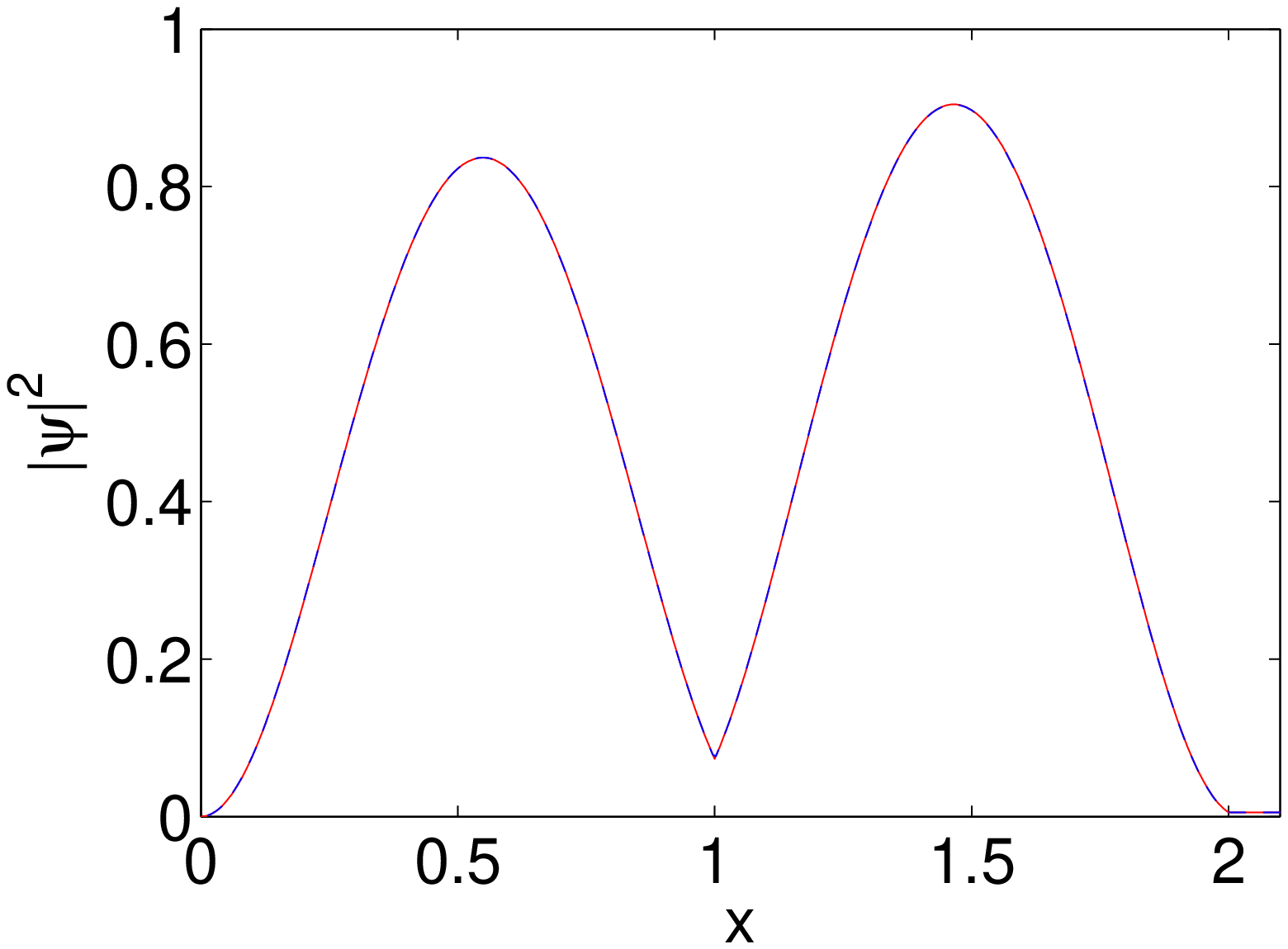}
\caption{\label{fig-DD_CAPS} {Comparison between the squared magnitudes of the analytically calculated
eigenstates (solid red) and the numerically exact solution obtained by the CAP method (dashed blue) for
the same states as in figure~\ref{fig-DD2}, i.e.~upper left: {autochtonous} self-trapping state (AuT),
lower left: {allochtonous} self-trapping state (AlT), upper right: allochtonous almost antisymmetric
state (Al-), lower right: autochtonous almost symmetric state (Au+). On the scale of drawing the results
of the two methods are almost indistinguishable.}} 
\end{figure}
\begin{table}[htbp]
\centering
\begin{tabular}{l|cc|cc}
 Eigenstate & $\mu_{\rm A}$ & $\Gamma_{\rm A}/2$ & $\mu_{\rm CAP}$ & $\Gamma_{\rm CAP}/2$ \\
  \hline
  {autochtonous} self-trapping state (AuT)       & 8.589 & 0.00114 & 8.588 & 0.00114   \\
  {allochtonous} self-trapping state (AlT)        & 8.301 & 0.02453 & 8.303 & 0.02443    \\
  allochtonous almost antisymmetric state (Al-)  & 6.999 & 0.01279 &       &   \\
  autochtonous almost symmetric state (Au+)      & 6.014 & 0.00902 & 6.014 & 0.00900    \\
\end{tabular}
\caption{Chemical potential and decay rates for the same states as in figure~\ref{fig-DD_CAPS},
The approximate analytical values (A) are compared with numerically exact ones (CAP)
calculated by a grid relaxation method with complex absorbing potentials.}
\label{tab-DDShell}
\end{table}
\begin{figure}[htb]
\centering
\includegraphics[width=7cm,  angle=0]{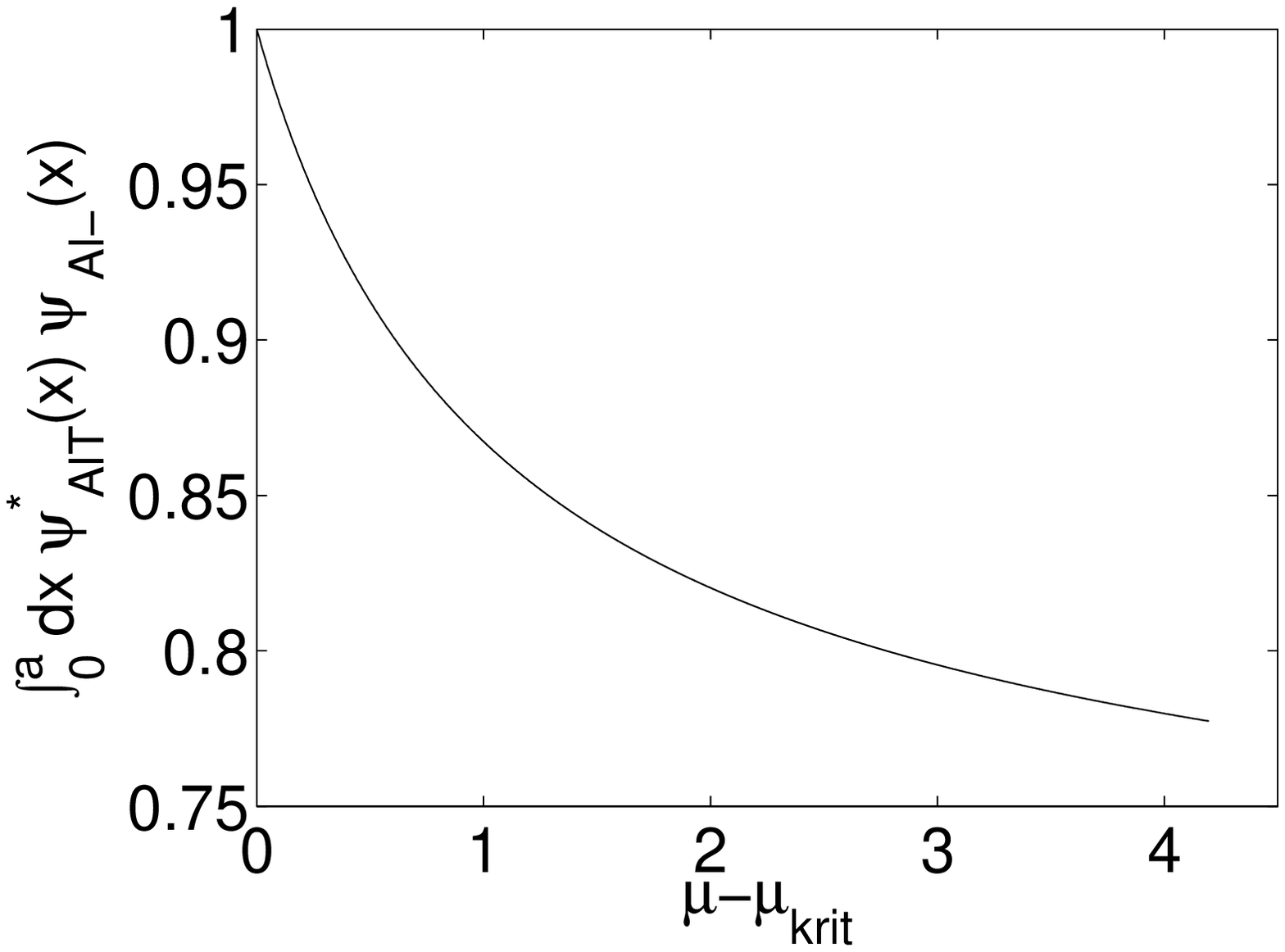}
\caption{\label{fig-DD_overlap} {Overlap of the two allochtonous states for the parameters
$\lambda_b=10$, $\lambda_a=20$, $b=1$, $a=2$.}}
\end{figure}

As an example we consider the lowest eigenstates of the potential with parameters $\lambda_b=10$, 
$\lambda_a=20$, $b=1$ and $a=2$.
Figure~\ref{fig-DD1} shows the real and imaginary parts of the eigenvalues in dependence of
the nonlinear parameter $g$. The
two lowest eigenvalues  $\mu$ of the linear ($g=0$) system (dashed green and dashed red) increase almost
linearly with increasing repulsive interaction strength $g$.
The corresponding wavefunctions are shown in figures~\ref{fig-DD_g0_5_g0} and \ref{fig-DD2} for $g=0$, $0.5$
and $3$. 
For a weak nonlinearity (figure~\ref{fig-DD_g0_5_g0}) the
the ground state is almost symmetric and the first excited state almost antisymmetric.
These states with linear counterpart are referred to as {\it autochtonous\/} states.
At $g \approx 1$ two new eigenvalues appear through a saddle node bifurcation (cf.~\cite{Theo06})
which we will henceforth call {\it allochtonous\/} states. 
One of these two states is dynamically stable 
(dashed dotted black) the other unstable (solid blue) (see subsection~\ref{subsec-DD_dynamics}). 
After the bifurcation the state, which was almost antisymmetric before the bifurcation,
(dashed green)
more and more localizes in the left well with increasing interaction, whereas the newly created
state (dashed dotted black) localizes in the right well (see upper left and lower left panel of figure~\ref{fig-DD2}).
These two symmetry breaking states are referred to as autochtonous self-trapping states (AuT) and
allochtonous self-trapping states (AlT), respectively. The two remaining states are referred to as
the {autochtonous} almost symmetric state (Au+) (dashed red) and {allochtonous} almost antisymmetric
state (Al-) (solid blue) (see upper right and lower right panel of figure~\ref{fig-DD2}).

The imaginary parts of the eigenvalues, the decay widths $\Gamma/2$, shown in the right panel of \ref{fig-DD1}
grow with increasing interaction strength $g$ for large values of $g$.
This behavior can be understood via the Siegert 
formula (\ref{DD-Siegert}) which predicts a dependence $\Gamma/2 \propto \sqrt{\mu}$ if the shape of 
the wavefunction does not change much with $\mu$ (respectively $g$) and furthermore $\mu$ is proportional
to $g$ in this parameter range (cf.~left panel of figure~\ref{fig-DD1}). For small values of $g$ the decay coefficient of the {autochtonous} self-trapping state (dashed green) decreases rapidly with increasing $g$
because of the increasing localization of the wavefunction in the left well where the probability for 
tunneling out of the barrier region $0 \le x \le a$ is small.

For comparison we consider the limit $\lambda_a \rightarrow \infty$ in which the system becomes both
symmetric and hermitian. Similar systems have recently been considered in a number of papers 
\cite{Infe06,Theo06,Khom07,Li06}.
Naturally the analytical results obtained in this limit are exact. 
From equation (\ref{DD-delta}) we see that in this case we have $\delta=0$ and all eigenvalues are real.
In the upper left panel of figure~\ref{fig-DD_u} the chemical potential $\mu$ is plotted in dependence 
of the interaction strength $g$. Compared to the nonhermitian nonsymmetric case considered before the
saddle node bifurcation has changed to a pitchfork bifurcation since the eigenvalues of the two 
self-trapping states now coincide due to the symmetry of the system (cf. \cite{Theo06}). One of
these states is shown in the lower left panel of figure~\ref{fig-DD_u}, the other one is obtained by
mirroring at the axis $x=b$. Hence the respective eigenvalues, indicated by the dashed dotted black
curve, are degenerate after the bifurcation. The two remaining eigenstates (lower right and upper 
right panel) are now genuinely symmetric, respectively antisymmetric.

As in the case of the open single well we compare our approximate analytical results for the resonance
solutions of the open double well with numerically exact ones calculated using a grid relaxation method with complex absorbing potentials
(see section~\ref{sec-single}). The
results are given in figure~\ref{fig-DD_CAPS} and table~\ref{tab-DDShell}.
For the two autochtonous states we observe good agreement between the analytical approximation and the
numerically exact solutions. For the AlT state (dashed dotted black) the grid relaxation only converges
for high values of $g$ (respectively $\mu$) but wherever it converges there is good agreement between 
the analytical approximation and the numerically exact solutions.
For the Al- state (solid blue) the grid relaxation does not converge at all. It turns out
(see sections \ref{subsec-DD_dynamics} and \ref{subsec-DD_BdG}) that this state is dynamically unstable
just like the respective state in the asymmetric hermitian double well (see \cite{Theo06}).

At the bifurcation point the real and imaginary parts of the eigenvalues of the {allochtonous} states
both coincide. By computing the overlap of the two respective wavefunctions (figure~\ref{fig-DD_overlap}) 
one can demonstrate that the wavefunctions also coincide at the bifurcation point. Thus the bifurcation
point is an example of an {\it exceptional\/} point (see \cite{03delta2,08PT,Cart08}).

\subsection{Dynamics}
\label{subsec-DD_dynamics}
The eigenstates calculated in the preceding subsection~have complex eigenvalues and are 
hence subject to decay. Following \cite{Schl06a,Schl06b} we call an eigenstate {\it dynamically stable\/}
if its time evolution follows the stationary adiabatic decay behavior given by 
\be
    \partial_t(g_{\rm eff})=-\Gamma(g_{\rm eff})\, g_{\rm eff} \label{DD-gN_dot}
\ee
where  $g_{\rm eff}=gN$ with $N=\int_0^a |\psi(x)|^2 \rd x$ denotes the effective nonlinear interaction in
inside the double well potential \cite{04nls_delta}.
In subsection~\ref{subsec-DD_stationary} the total norm was kept fixed at $N=1$ and $g$ was varied.
Now we keep $g$ fixed whereas $N$ decreases due to decay.
We compute the time evolution of the states shown in figure \ref{fig-DD2} using a Crank-Nicholson
propagation with a predictor-corrector algorithm and absorbing boundaries \cite{Paul07b,Shib91}.

\begin{figure}[htb]
\centering
\begin{minipage}{7cm}
\includegraphics[width=7cm,  angle=0]{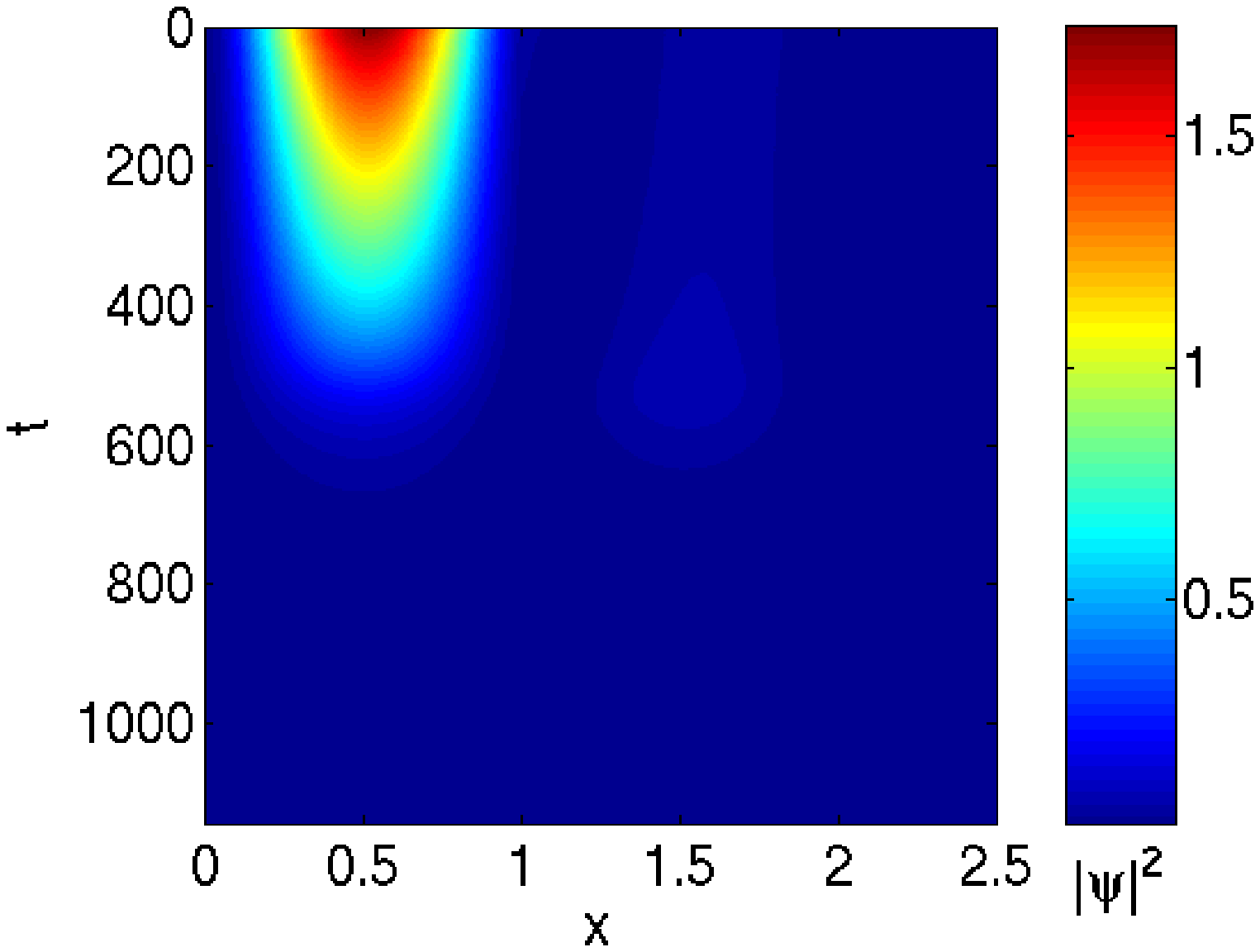}
\end{minipage}
\begin{minipage}{7cm}
\includegraphics[width=7cm,  angle=0]{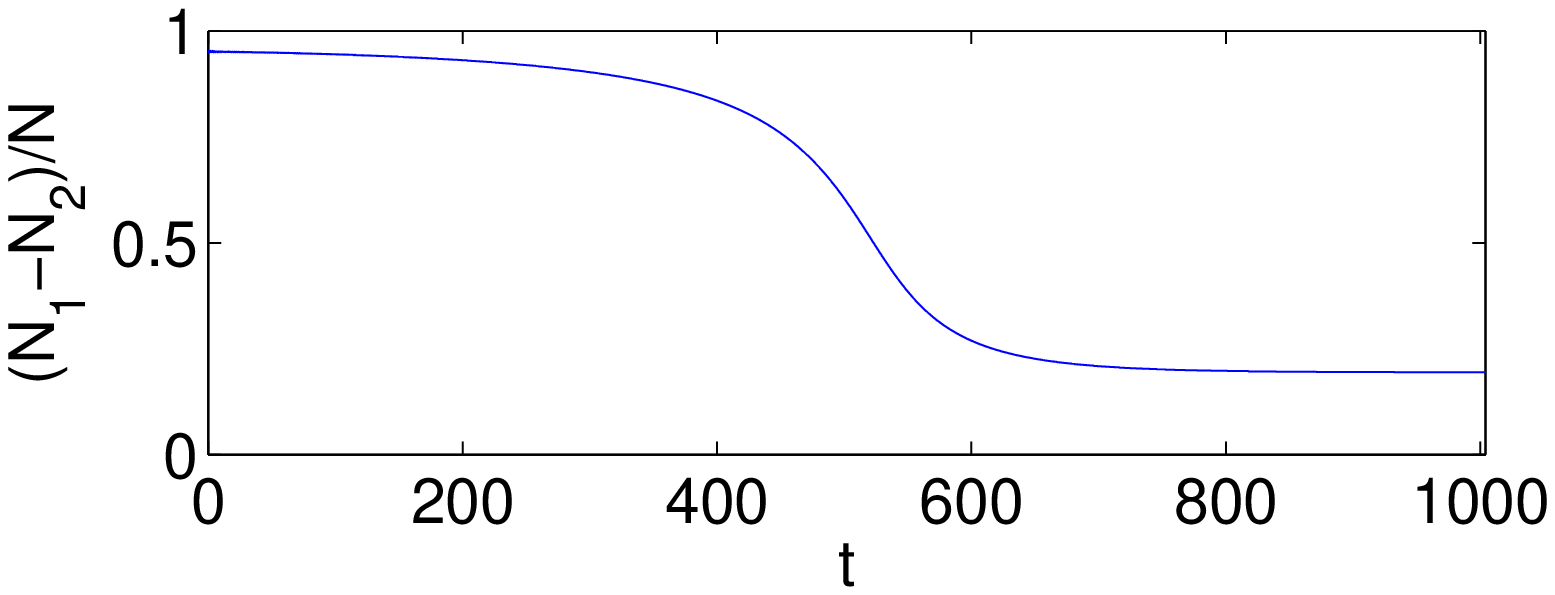}\\
\includegraphics[width=7cm,  angle=0]{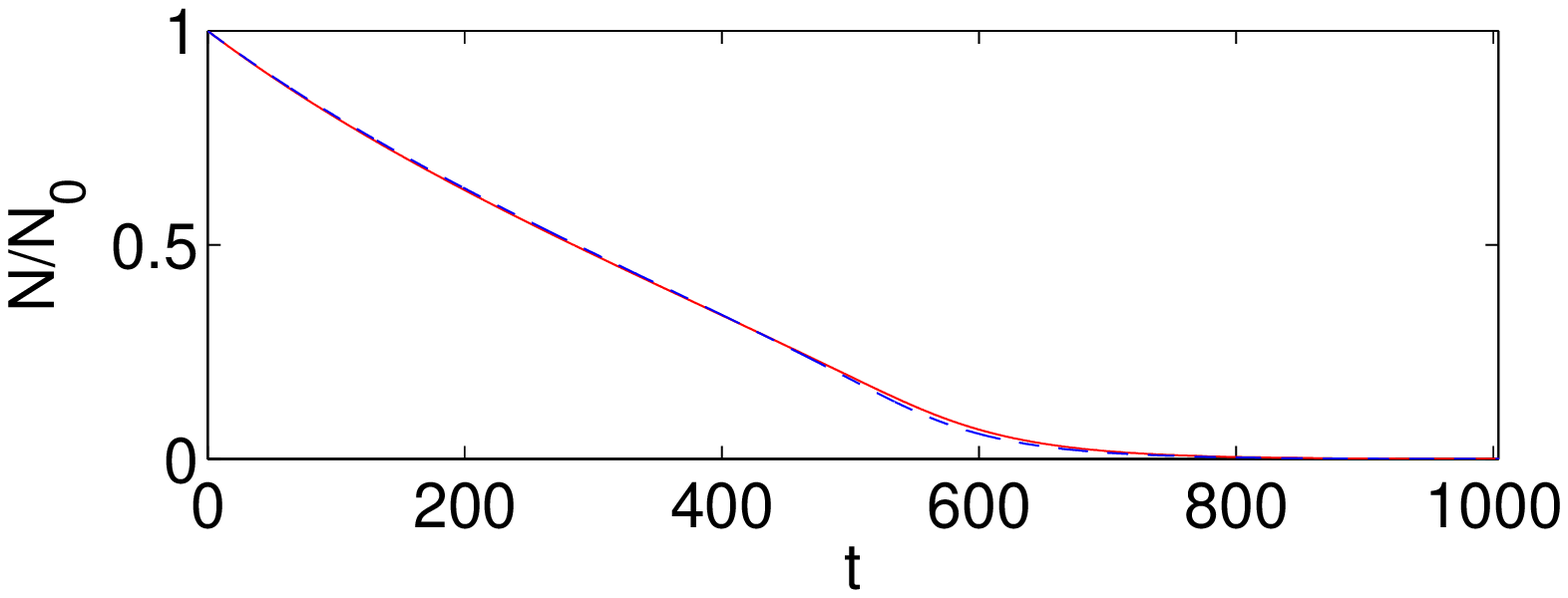}
\end{minipage}
\caption{\label{fig-DD_Dyn1} {Time evolution of the {autochtonous} self-trapping state (AuT) from
figure~\ref{fig-DD2}. Left panel: Spatio-temporal contour plot of the density $|\psi(x)|^2$.
Upper right panel: relative population of the two wells. $N_1$ and $N_2$ denote the total norm inside
the left respectively right well. Lower right panel: Decay of the norm $N$ of the wavefunction inside 
the double barrier calculated via time evolution (dashed blue) and using stationary states (solid red).
$N_0$ is the norm at $t=0$.}} 
\end{figure}
\begin{figure}[htb]
\centering
\begin{minipage}{7cm}
\includegraphics[width=7cm,  angle=0]{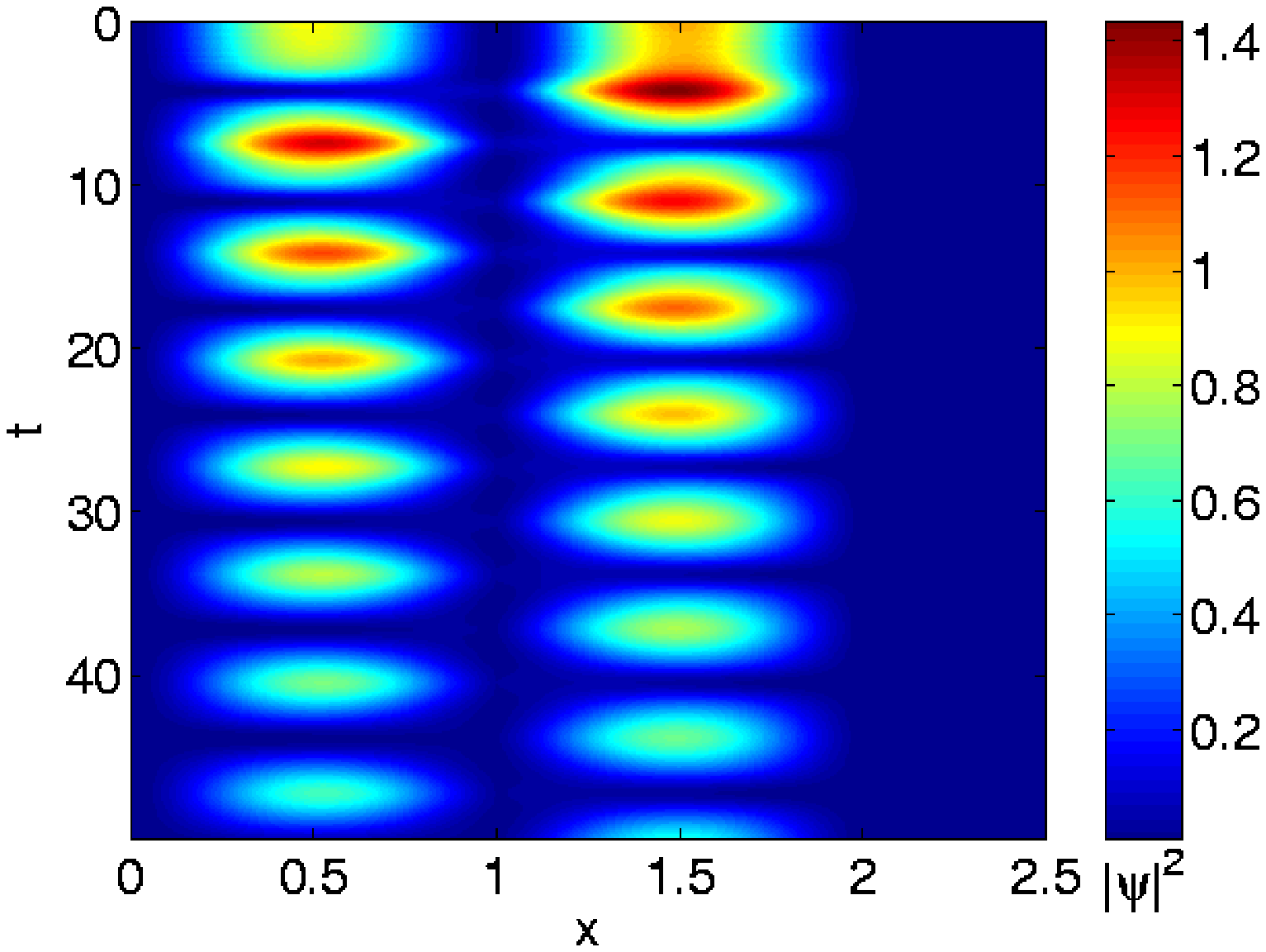}
\end{minipage}
\begin{minipage}{7cm}
\includegraphics[width=7cm,  angle=0]{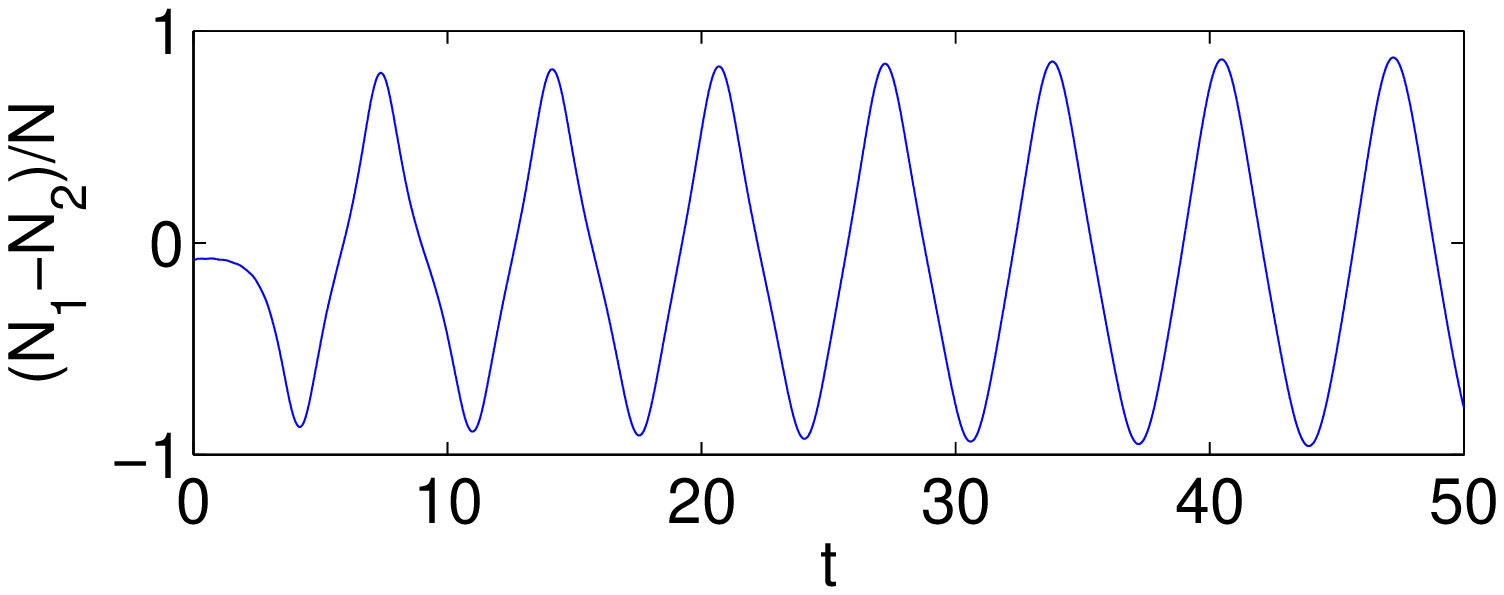}\\
\includegraphics[width=7cm,  angle=0]{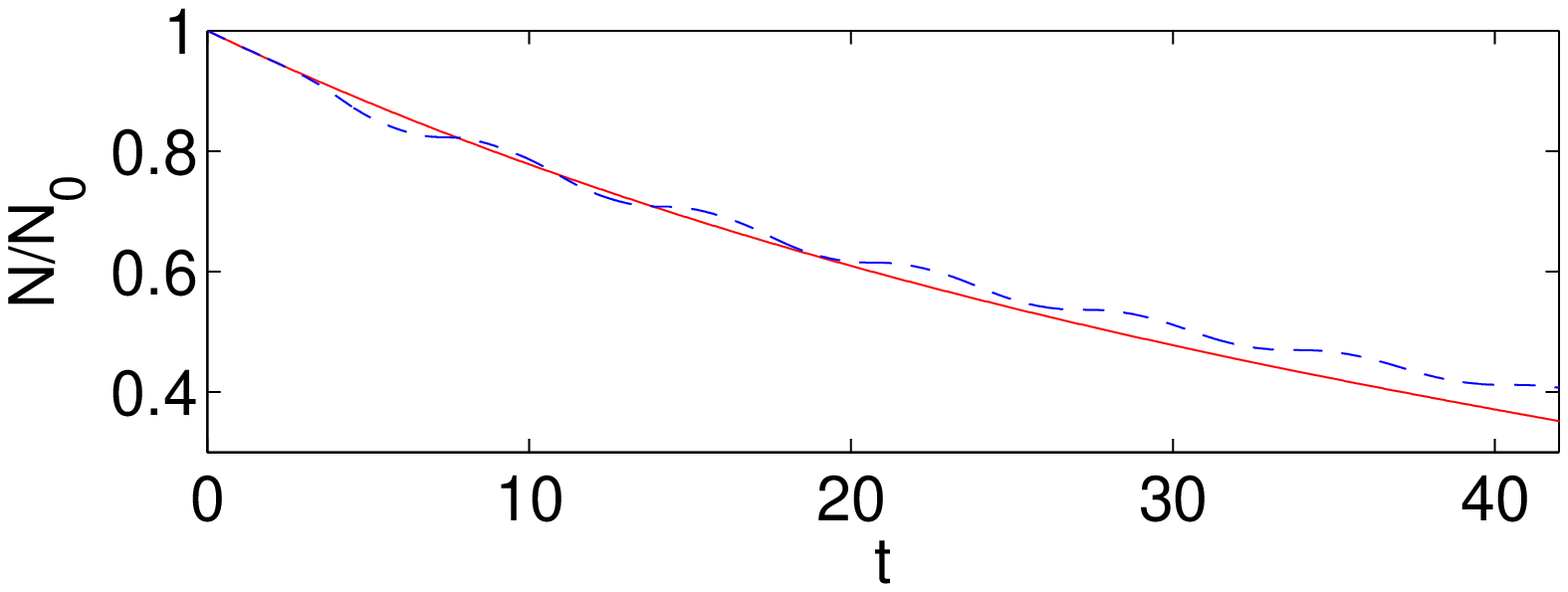}
\end{minipage}
\caption{\label{fig-DD_Dyn2} {The same as figure \ref{fig-DD_Dyn1}, however, for the allochtonous almost antisymmetric state (Al-).
}} 
\end{figure}
\begin{figure}[htb]
\centering
\begin{minipage}{7cm}
\includegraphics[width=7cm,  angle=0]{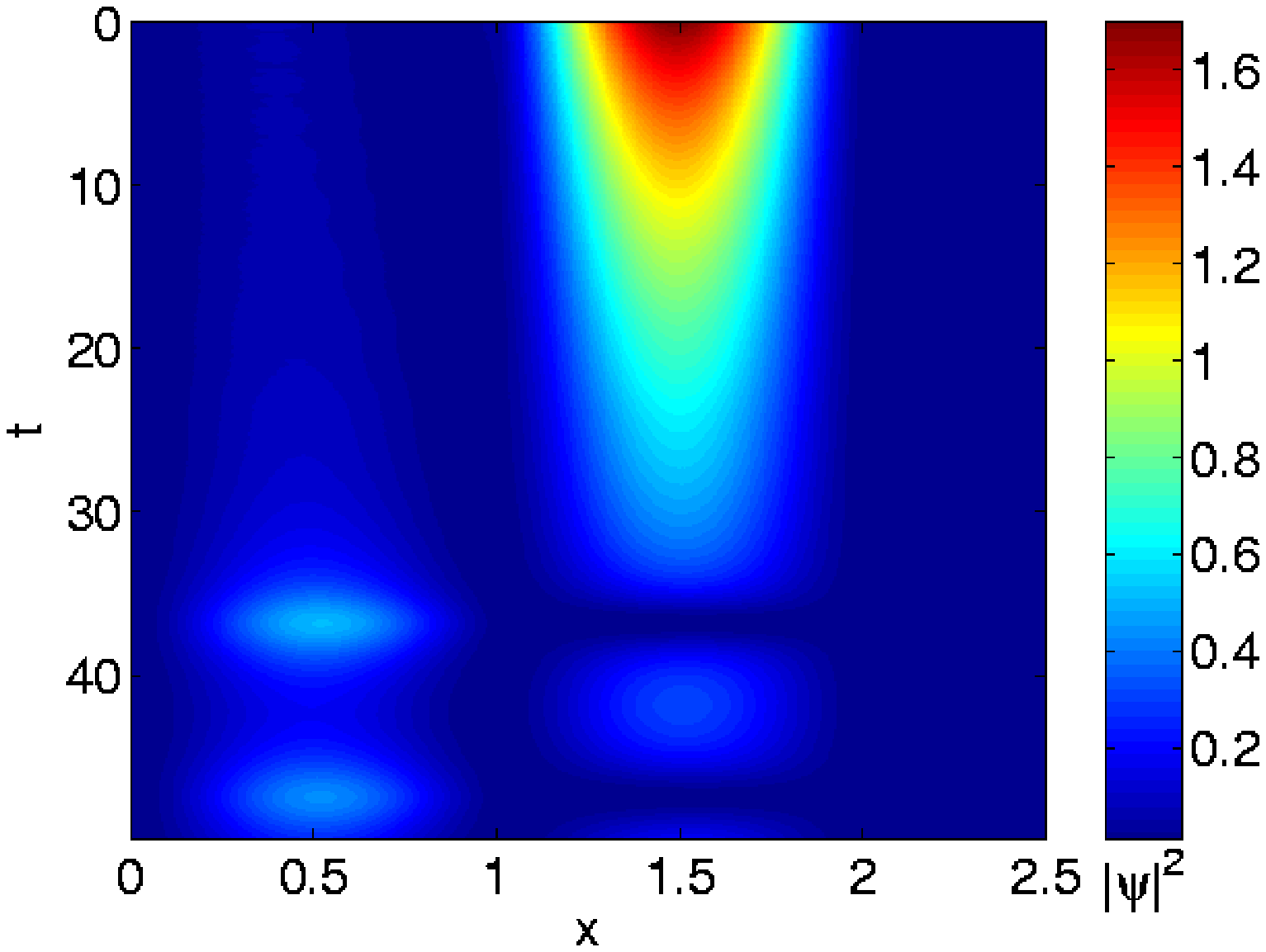}
\end{minipage}
\begin{minipage}{7cm}
\includegraphics[width=7cm,  angle=0]{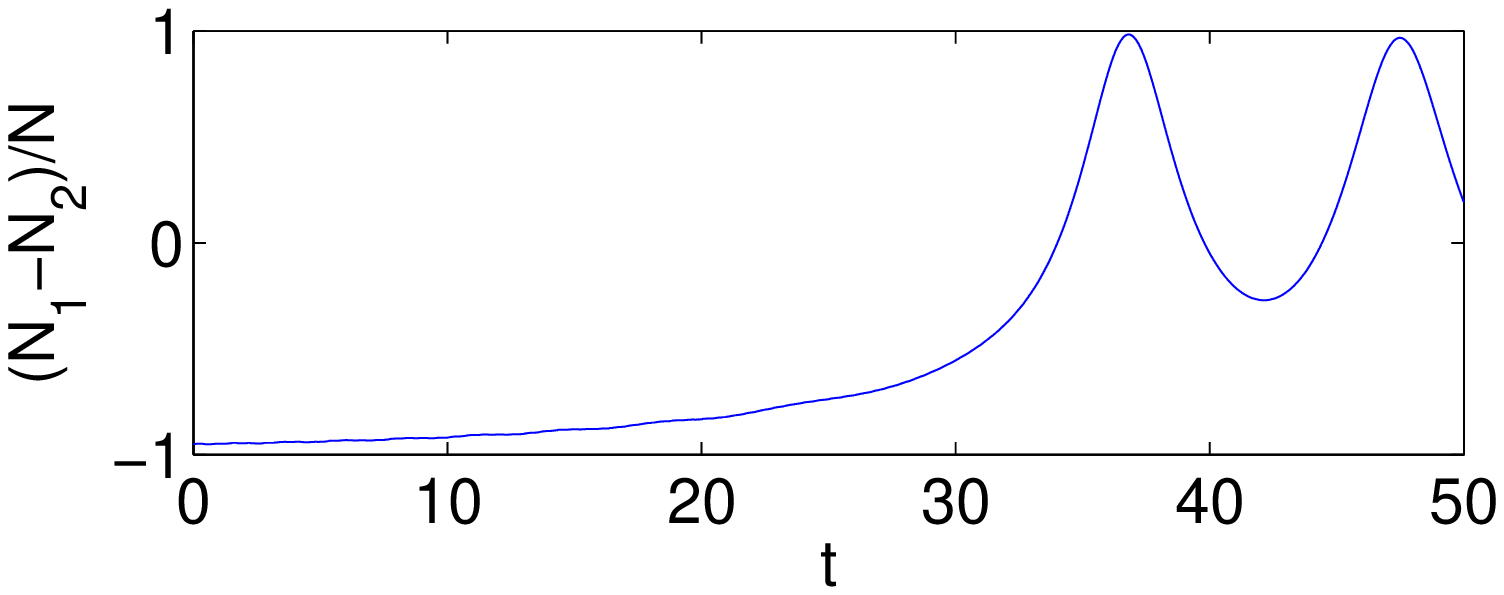}\\
\includegraphics[width=7cm,  angle=0]{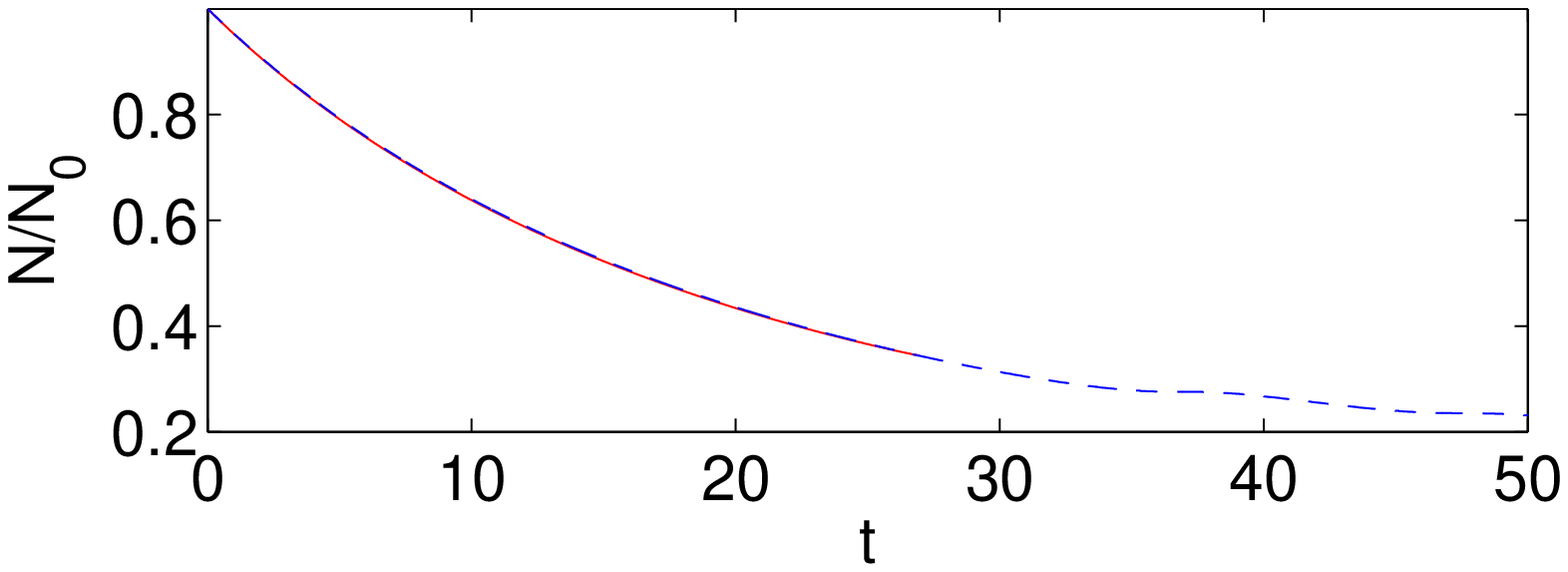}
\end{minipage}
\caption{\label{fig-DD_Dyn3} {The same as figure \ref{fig-DD_Dyn1}, however, for the allochtonous self-trapping state (AlT).
}} 
\end{figure}
\begin{figure}[htb]
\centering
\begin{minipage}{7cm}
\includegraphics[width=7cm,  angle=0]{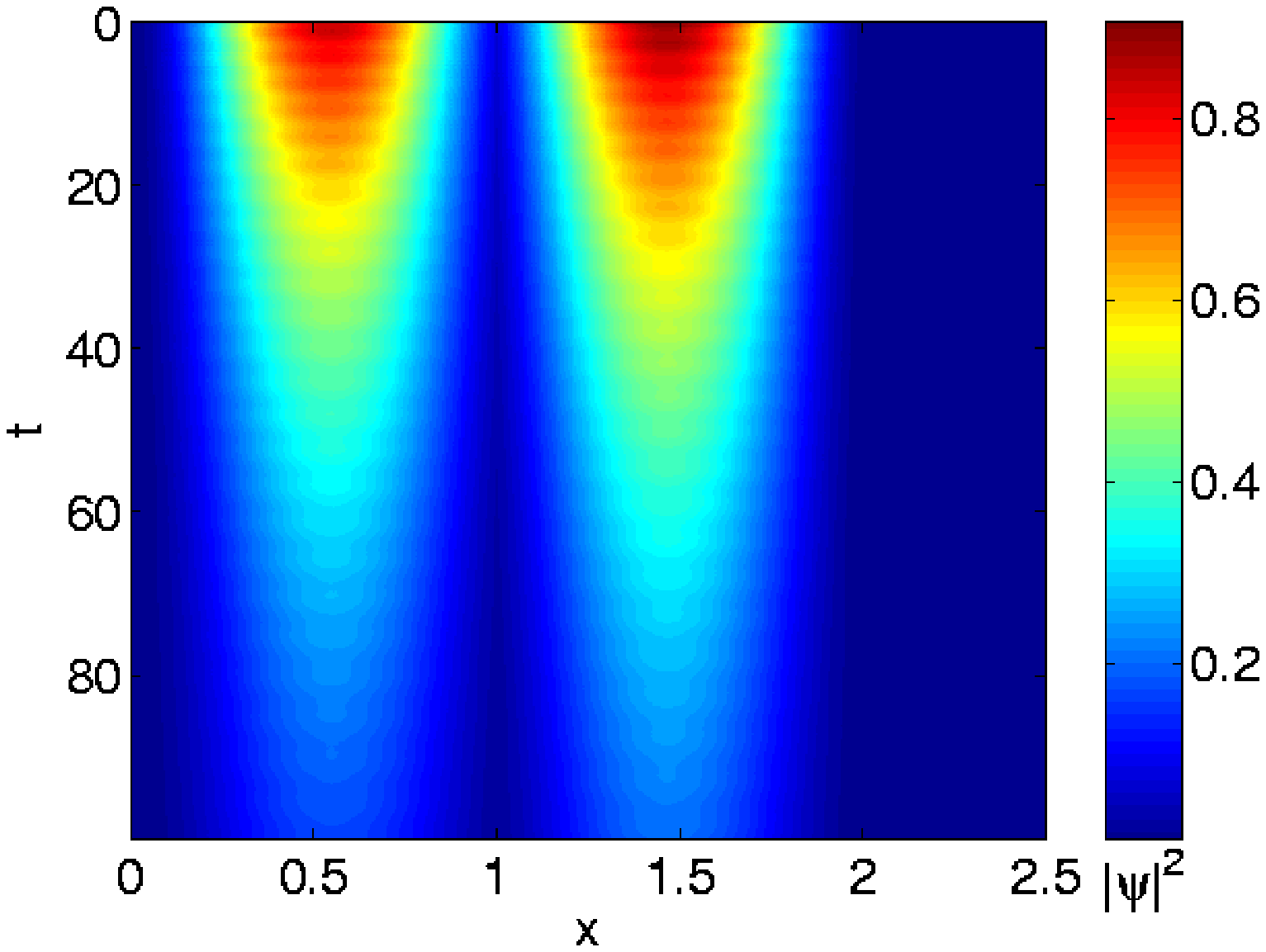}
\end{minipage}
\begin{minipage}{7cm}
\includegraphics[width=7cm,  angle=0]{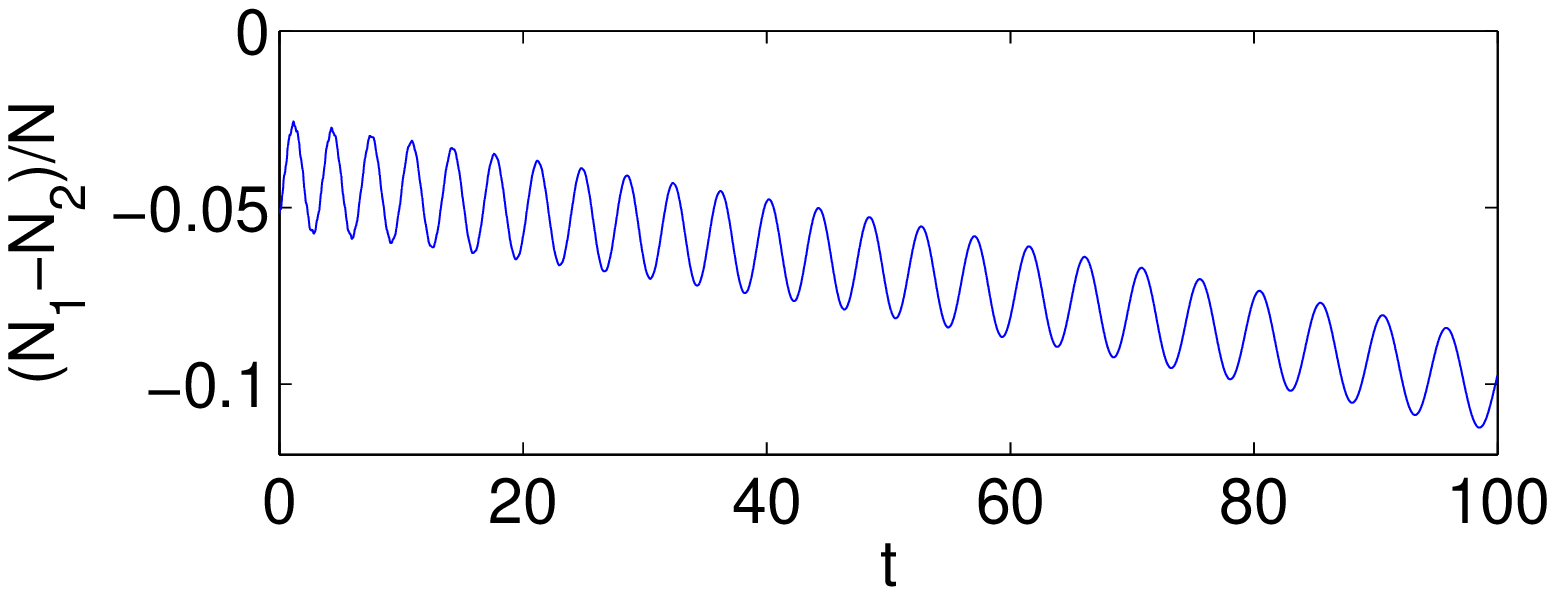}\\
\includegraphics[width=7cm,  angle=0]{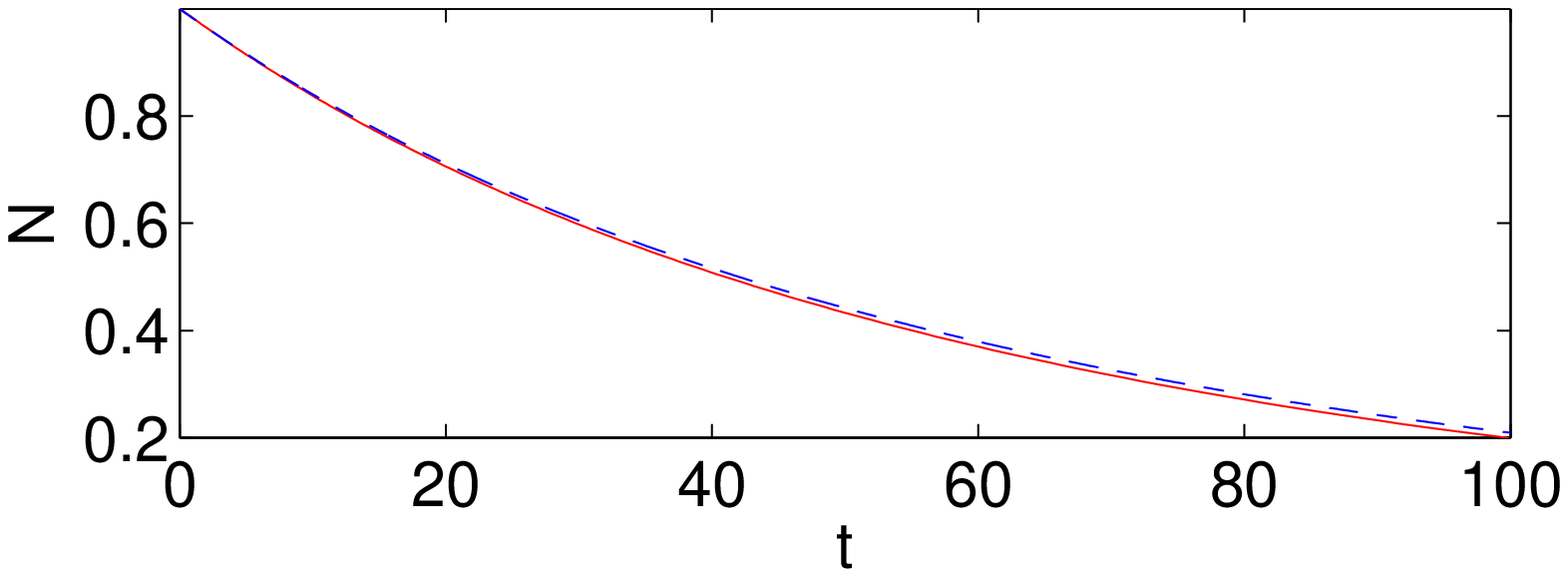}
\end{minipage}
\caption{\label{fig-DD_Dyn4} {The same as figure \ref{fig-DD_Dyn1}, however, for the autochtonous almost symmetric state (Au+).
}} 
\end{figure}

The results are given in figures \ref{fig-DD_Dyn1} -- \ref{fig-DD_Dyn4} each of  which shows a spatio-temporal 
contour plot of the density $|\psi(x,t)|^2$ (left panel), the relative population of the two wells
(upper right panel) and the decay of the norm $N$ (lower right panel) which is calculated directly via 
time-evolution (dashed blue curve) and compared to the stationary decay behavior according to equation (\ref{DD-gN_dot})
(solid red curve) calculated from the stationary eigenvalues shown in figure \ref{fig-DD1}.
The time-evolution of the {autochtonous} states (figures \ref{fig-DD_Dyn1} and \ref{fig-DD_Dyn4}) closely
follows the stationary decay behavior. Thus these states are dynamically stable. The relative population of the state Au+ (upper right panel in figure \ref{fig-DD_Dyn4}) reveals a small oscillation in addition to the adiabatic evolution which can be explained by a linear stability analysis based on the Bogoliubov-de-Gennes equations (see section \ref{subsec-DD_BdG}).

If the time-evolution is initiated with the almost antisymmetric {allochtonous} state 
(Al-, figure~\ref{fig-DD_Dyn2}) 
the wavefunction starts to oscillate between the wells and hence does not exhibit a stationary decay behavior so that it is termed dynamically unstable. Due to the asymmetry of the system there is no complete population transfer between the wells. Because of the influence of interaction the Josephson oscillations for small times are not sine-shaped but can be described by a Jacobi elliptic function (see e.g.~\cite{Khom07} and references therein). For longer times the effective nonlinearity decreases due to decay so that the oscillation becomes more sine-shaped and its period slowly gets closer to the value $2\pi \hbar/\Delta\mu_{g=0}\approx 7.53$ for the Josephson oscillations of the linear system where $\Delta\mu_{g=0}\approx 0.834$ is the difference between the chemical potentials of the ground and first excited state for $g=0$ (cf.~figure \ref{fig-DD_g0_5_g0}). In order to estimate the oscillation period for short times we consider the interval between the first two population maxima in the left well at $t_1\approx 4.05$ and $t_2 \approx 10.84$. In an ad hoc approach we take the interaction into account by considering the difference between the chemical potentials of the almost symmetric, respectively antisymmetric quasistationary eigenstates Au+ and Al- in the middle of the interval $[t_1,t_2]$ instead of $\Delta\mu_{g=0}$.
In the middle of the interval $[t_1,t_2]$ we still have about $80\%$ of the initial population left so that the effective nonlinearity can be estimated as $g_{\rm eff} \approx 2.4$ and the respective difference between the chemical potentials is approximately given by $\Delta\mu_{g=2.4}\approx 0.95$ (cf.~figure \ref{fig-DD1}). The resulting estimate for the period $2 \pi \hbar/\Delta\mu_{g=2.4} \approx 6.6$ roughly agrees with the value $t_2-t_1 \approx 6.79$ observed in the dynamical calculation.

The time-evolution of the {allochtonous} self-trapping state (AlT, figure~\ref{fig-DD_Dyn3}) follows the
stationary decay behavior until the bifurcation point is reached. After that it tunnels completely from the right to the left well and starts oscillating between total population of the left well and an intermediate population of the two wells. This asymmetry indicates that a part of the total population of the double well does not take part in the oscillation but is trapped in the AuT state.

In conclusion, we observe that there are dynamically stable states whose dynamics is well described by 
stationary states in an adiabatic approximation. In particular, the self-trapping effect is not
immediately destroyed by decay but survives until the bifurcation point is reached. Especially in the 
{autochtonous} self-trapping state (AuT) the trapping can be preserved a long time since this state 
decays very slowly.


\subsection{Finite basis approximation}
\label{subsec-DD_Galerkin}
Instead of solving the NLSE as a differential equation, one can,
in a Galerkin-type ansatz, expand the wavefunction $\psi(x,t)=\exp(-\ri (\mu-\ri \Gamma/2) t/\hbar)u(x,t) $ as
\be
   u(x,t)=\sum_{j=1}^{n_{\rm B}} c_j(t)\, u_j(x)
\ee
using the first $n_{\rm B}$ eigenfunctions $\{ u_j \}$ and respective  eigenvalues 
$\{ \mu_j-\ri \Gamma_j/2\}$ 
of the corresponding linear ($g=0$) system with the Hamiltonian
\be
H_0=-\frac{\hbar^2}{2m} \partial_x^2 +V(x)\,
   \label{H0_u}
\ee
using Siegert boundary conditions.
The NLSE can now be written as
\be
   \ri \hbar \partial_t u= \Big(H_0-(\mu-\ri \Gamma/2)\Big)\,u +g|u|^2u\, .
   \label{GPE_u}
\ee
Since $H_0$ is not hermitian the states $\{ u_j \}$ are not orthogonal.
Instead they form a \emph{finite} basis set
\be
   \int_{0}^{\infty} v_j^*(x)\, u_l(x) \rd x =\delta_{jl}
\ee
together with the eigenvectors $\{v_j\}$ of the adjoint Hamiltonian $H_0^\dagger$ satisfying
\be
   H_0^\dagger\, v_j=(\mu_j-\ri \Gamma_j/2)^*\, v_j\,,
   \label{DDShell_H0Dagger}
\ee
the so-called the \emph{left eigenvectors} of $H_0$ (see e.g. \cite{Mors53}). We compute the left and 
right eigenvectors using
exterior complex scaling in order to make the states $\{u_j\}$ and $\{v_j\}$ square integrable in $0 \le x \le \infty$ (see \ref{app_DDShell_lin}).

To calculate the stationary states we set $\partial_t u_j=0$, $1 \le j \le n_{\rm B}$, and consider the projections
of equation (\ref{GPE_u}) on the $n_{\rm B}$ left eigenvectors:
\be \fl
  c_j\,\big(E_j -E\big) + g \int_{0}^{\infty} \rd x \, v_j^*(x) \left| \sum_{i=1}^{n_{\rm B}} c_i u_i(x)
  \right|^2 \sum_{l=1}^{n_{\rm B}} c_l u_l(x) =0 \ , \quad 1 \le j \le n_{\rm B} \label{DD-u_j}
\ee
with $E=\mu-\ri \Gamma/2$ and $E_j=\mu_j-\ri \Gamma_j/2$ .
Together with the normalization condition
\be
   \int_0^a \rd x \,  |u(x)|^2 = 1
\ee
and a condition
\be
   \arg(u(a))=0 \label{DD-arg}
\ee
which fixes the phase of the wavefunction we obtain a system of nonlinear equations for the coefficients $\{ c_j \}$,
the chemical potential $\mu$ and the decay rate $\Gamma$. This system (\ref{DD-u_j}) -- (\ref{DD-arg}) is solved with a Newton 
algorithm.

Alternatively, equation (\ref{DD-u_j}) can be rewritten as
\be
  c_j\,\big(E_j -E\big) + \sum_{ik\ell =1}^{n_{\rm B}}
  w_{k\ell}^{ji}\,c_i^*\,c_k\,c_\ell=0
\ee
with
\be
 w_{k\ell}^{ji}=g\int_{0}^{\infty} \rd x \, v_j^*(x) u_i^*(x)u_k(x)u_\ell(x)=w_{\ell \, k}^{ji}
\ee
where the integrations can be carried out once in the beginning. This has some advantages for small $n_{\rm B}$.
For the most simple case of only two states this yields the nonlinear, nonhermitian $2\times 2$ eigenvalue
equation 
\be 
\begin{pmatrix}
E_1-E+\kappa_{11}&\kappa_{12}\\
\kappa_{21}& E_2-E+\kappa_{22}
\end{pmatrix}\begin{pmatrix}c_1\\c_2\end{pmatrix}
=0
\ee
with
\begin{eqnarray}
\kappa_{11}&=&w_{11}^{11}|c_1|^2 +2w_{12}^{12}|c_2|^2 +w_{11}^{12}c_2^*c_1\\
\kappa_{12}&=&w_{11}^{22}|c_2|^2 +2w_{11}^{12}|c_1|^2 +w_{11}^{22}c_1^*c_2\\
\kappa_{21}&=&w_{11}^{21}|c_1|^2 +2w_{12}^{21}|c_2|^2 +w_{11}^{21}c_2^*c_1\\
\kappa_{22}&=&w_{22}^{22}|c_2|^2 +2w_{12}^{21}|c_1|^2 +w_{22}^{21}c_1^*c_2 \, .
\end{eqnarray}

\begin{figure}[htb]
\centering
\includegraphics[width=7cm,  angle=0]{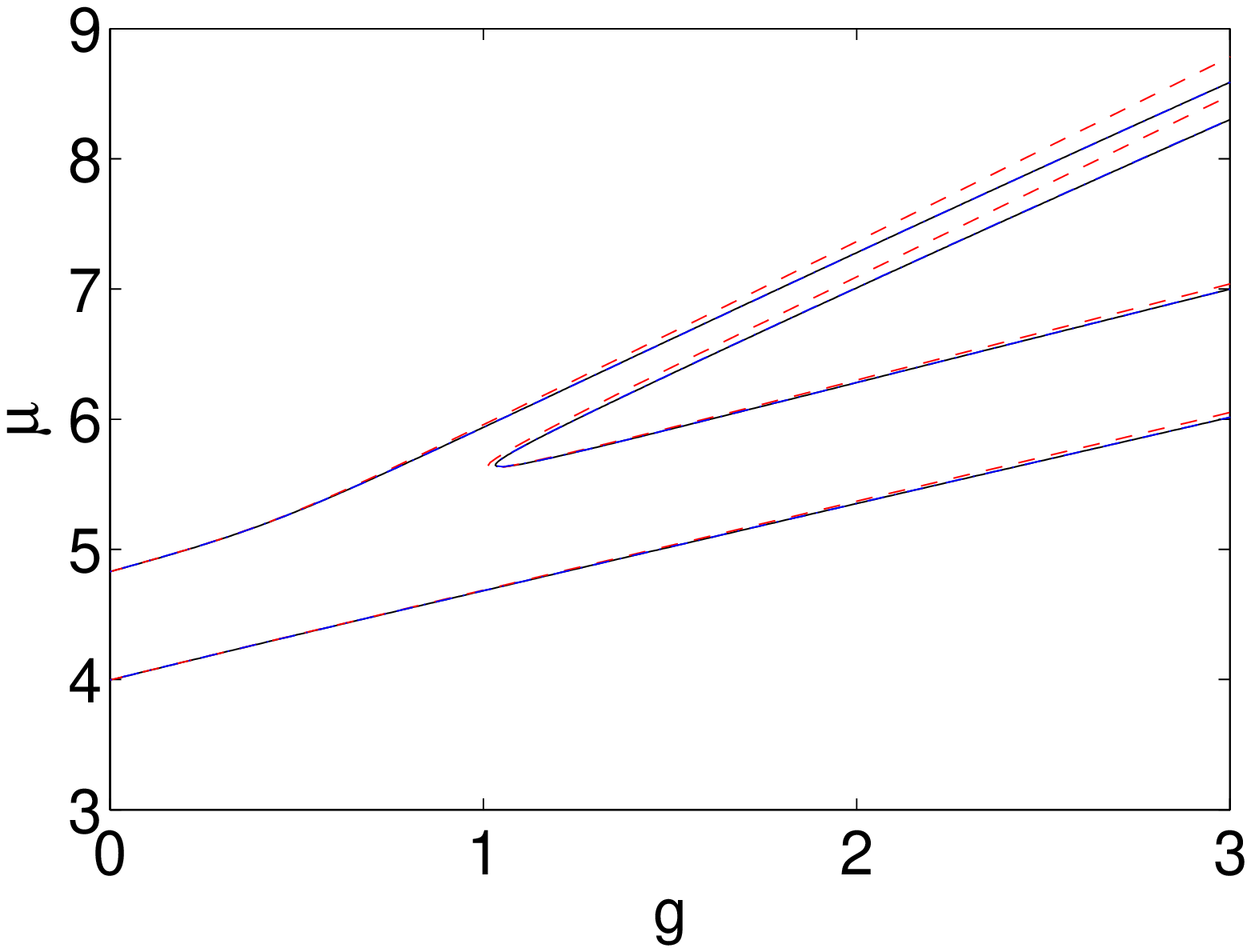}
\includegraphics[width=7cm,  angle=0]{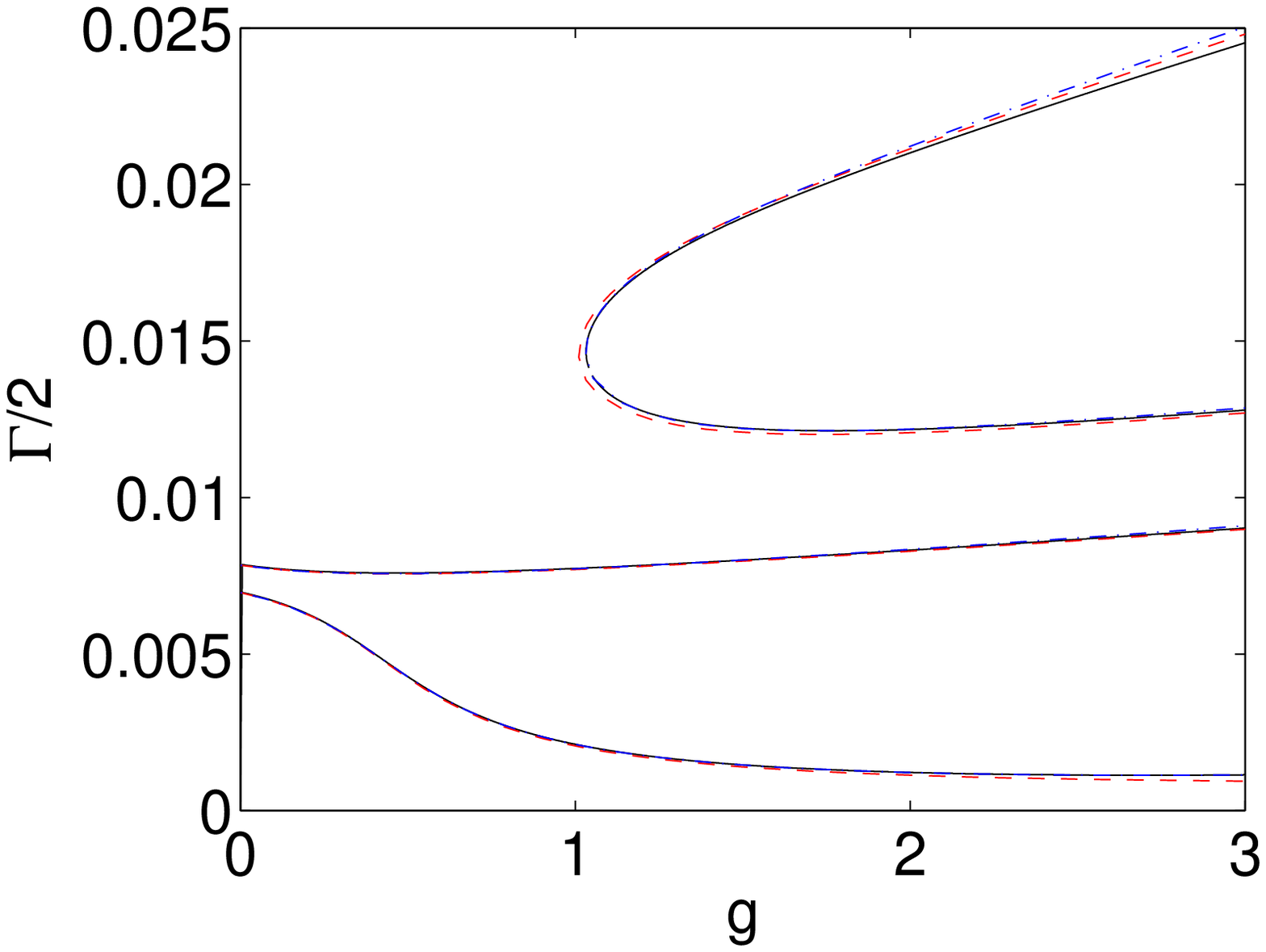}\\
\caption{\label{fig-DD_B} {Bifurcation diagram for $\lambda_b=10$, $\lambda_a=20$, $b=1$, $a=2$.
Solid black: analytical result (cf. figure~\ref{fig-DD1}), dashed red: Galerkin approximation with 
$n_{\rm B}=2$ modes, dashed dotted blue: $n_{\rm B}=30$ modes.}}
\end{figure}

In figure~\ref{fig-DD_B} we compare the bifurcation diagram calculated in section~\ref{subsec-DD_stationary} 
(solid black lines) with the Galerkin approach with $n_{\rm B}=2$, (dashed red lines) and $n_{\rm B}=30$ modes
(dashed dotted blue lines).
Both real (left panel) and imaginary (right panel) parts of the eigenvalues are reasonably well 
reproduced by such a two mode approximation. Naturally the agreement is best for the two lowest 
states. The Galerkin approximation with $n_{\rm B}=30$ modes almost coincides with the results 
from section~\ref{subsec-DD_stationary} in the displayed parameter range $0 \le g\le 3$ except 
for the decay coefficient of the {allochtonous} self-trapping state for high values of $g$.

\begin{figure}[htb]
\centering
\includegraphics[width=7cm,  angle=0]{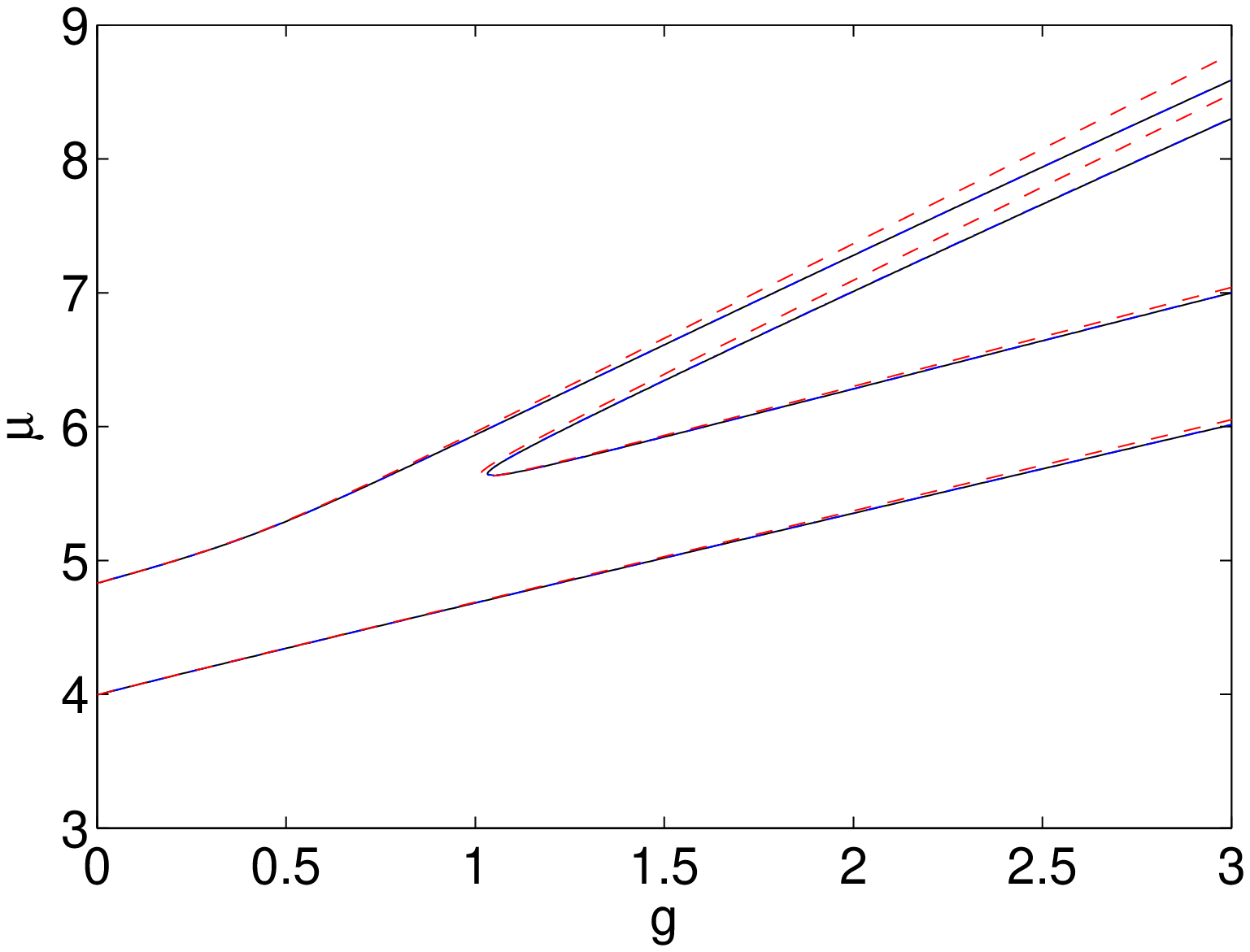}
\includegraphics[width=7cm,  angle=0]{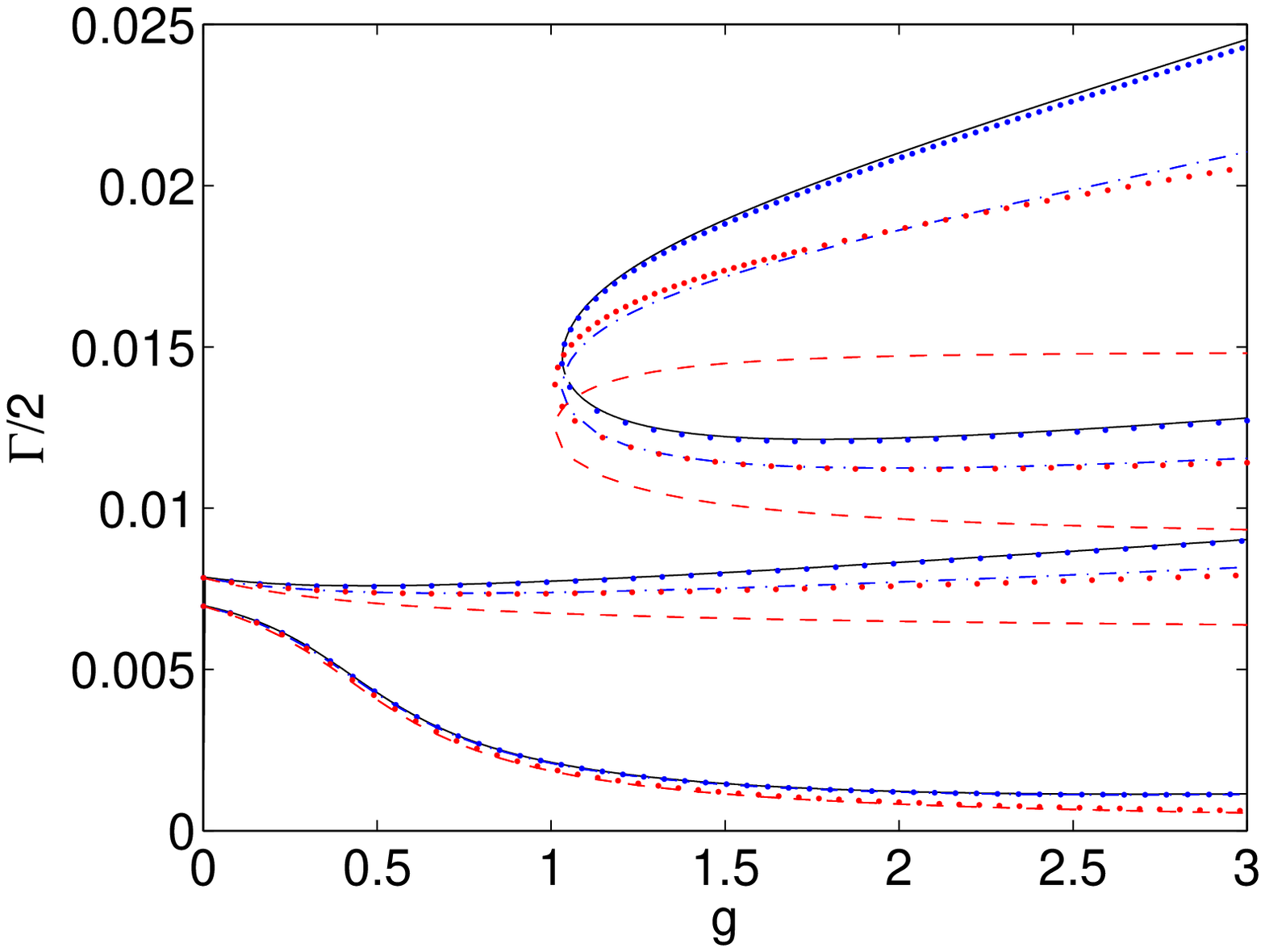}\\
\caption{\label{fig-DD_B_alt} {Bifurcation diagram for $\lambda_b=10$, $\lambda_a=20$, $b=1$, $a=2$ and projection onto the right eigenvectors $\{u_j\}$. Solid black: analytical result
(cf. figure~\ref{fig-DD1}), dashed red: Galerkin approximation with $n_{\rm B}=2$ modes, red dots:
$n_{\rm B}=2$ modes and Siegert formula, dashed dotted blue: $n_{\rm B}=10$ modes, blue dots: 
$n_{\rm B}=10$ modes and Siegert formula}}
\end{figure}
In order to achieve such a good agreement, it is essential to use the biorthogonal basis.
If only the  right eigenvectors are used, i.e.~if we
project equation (\ref{GPE_u}) on the $\{u_j\}$ instead of  the $\{v_j\}$, the results
are worse.
Figure~\ref{fig-DD_B_alt} reveals that for $n_{\rm B}=2$ modes the real parts of the eigenvalues 
are still in good agreement with the analytical results whereas the imaginary parts of the 
eigenvalues clearly differ from the previous results since the growth of the decay rates with 
increasing nonlinearity $g$ (cf.~the discussion in section~\ref{subsec-DD_stationary}) is not reproduced.
The results for $n_{\rm B}=10$ demonstrate that the agreement improves if higher modes are taken 
into account.  Thus the decay rates are much more sensitive to the excitation of higher modes than
the real parts of the eigenvalues. The agreement can be improved if we calculate the decay rates 
with the Siegert formula (\ref{DD-Siegert}) using the wavefunctions and real parts of the 
respective eigenvalues calculated with the $n_{\rm B}=2$ (red dots) and the $n_{\rm B}=10$ 
(blue dots) Galerkin approximation.

These results indicate that the  biorthogonal basis $\{v_j\}$, $\{u_j\}$ is better suited to 
describe the system than
$\{u_j\}$ alone. Naturally the differences between different choices of basis sets disappear 
in the limit $n_{\rm B} \rightarrow \infty$.

\subsection{Linear stability analysis}
\label{subsec-DD_BdG}
In this subsection~we analyze the stability of the adiabatic time evolution of the quasistationary 
states by solving the Bogoliubov-de-Gennes (BdG) equations and compare its predictions with the results 
of the dynamical calculations from section~\ref{subsec-DD_dynamics}. The BdG equations
are obtained by linearizing the GPE in the vicinity of a background solution $\psi_0$, i.e.~by inserting
$\psi=\psi_0+\delta \psi$ into the GPE and expanding the resulting equation up to first order in 
$\delta \psi$. Since we are considering quasistationary rather than stationary background solutions we
follow the prescription of Castin and Dum \cite{Cast98} for time-dependent background solutions. They 
argue that only the components $\delta \psi_\perp$ perpendicular to the time-dependent background
solution are relevant. Assuming a time-dependence
$\delta \psi(t)=\delta \psi \exp[-(\mu-\ri \Gamma/2)t/\hbar-\ri \omega t]$ the BdG equations read in our case
\be \fl
\hbar \omega \begin{pmatrix}\delta \psi_\perp\\\delta \psi_\perp^* \end{pmatrix}=
\begin{pmatrix}H_{\rm GP}+gQ|\psi_0|^2Q-E& g Q \psi_0^2Q^* \\ 
-gQ^*{\psi_0^*}^2&-H_{\rm GP}^*-gQ^*|\psi_0|^2Q^*+E^*\end{pmatrix}
\begin{pmatrix}\delta \psi_\perp\\\delta \psi_\perp^* \end{pmatrix}
\label{DD-BdG}
\ee
with $E=\mu-\ri \Gamma/2$, where
$Q=1-|\psi_0\rangle \langle \psi_0|$ is the projector orthogonal to the background state $\psi_0$ and the
Gross-Pitaevskii Hamiltonian $H_{\rm GP}=H_0+g|\psi_0|^2$. The population of eigenmodes with positive 
imaginary part of the eigenvalue grows exponentially in time, thus leading to instability.
The eigenvalue equation (\ref{DD-BdG}) is solved by expansion in the eigenbasis of $H_0$ as described in 
section~\ref{subsec-DD_Galerkin}.
\begin{figure}[htb]
\centering
\includegraphics[width=7cm,  angle=0]{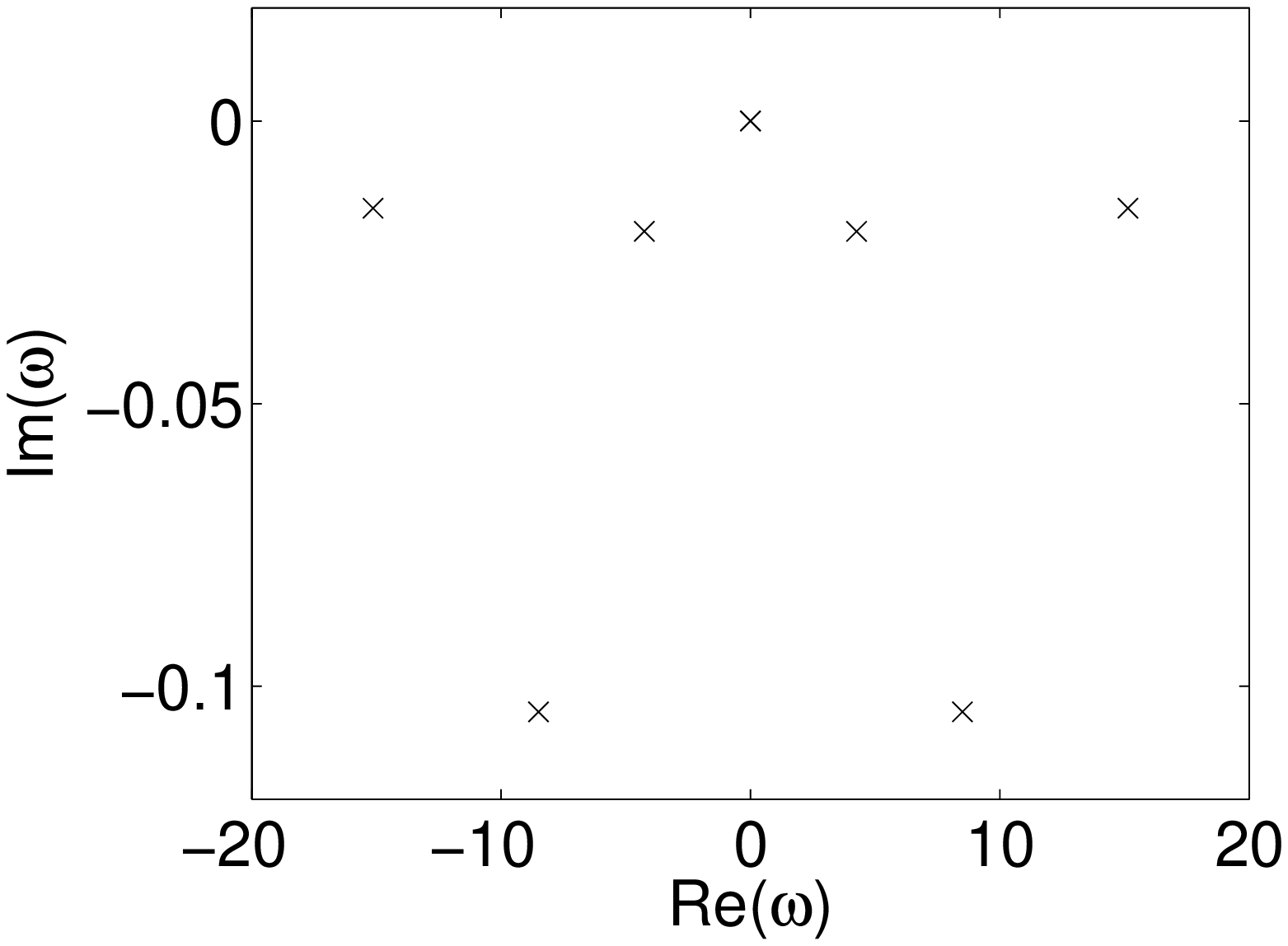}
\includegraphics[width=7cm,  angle=0]{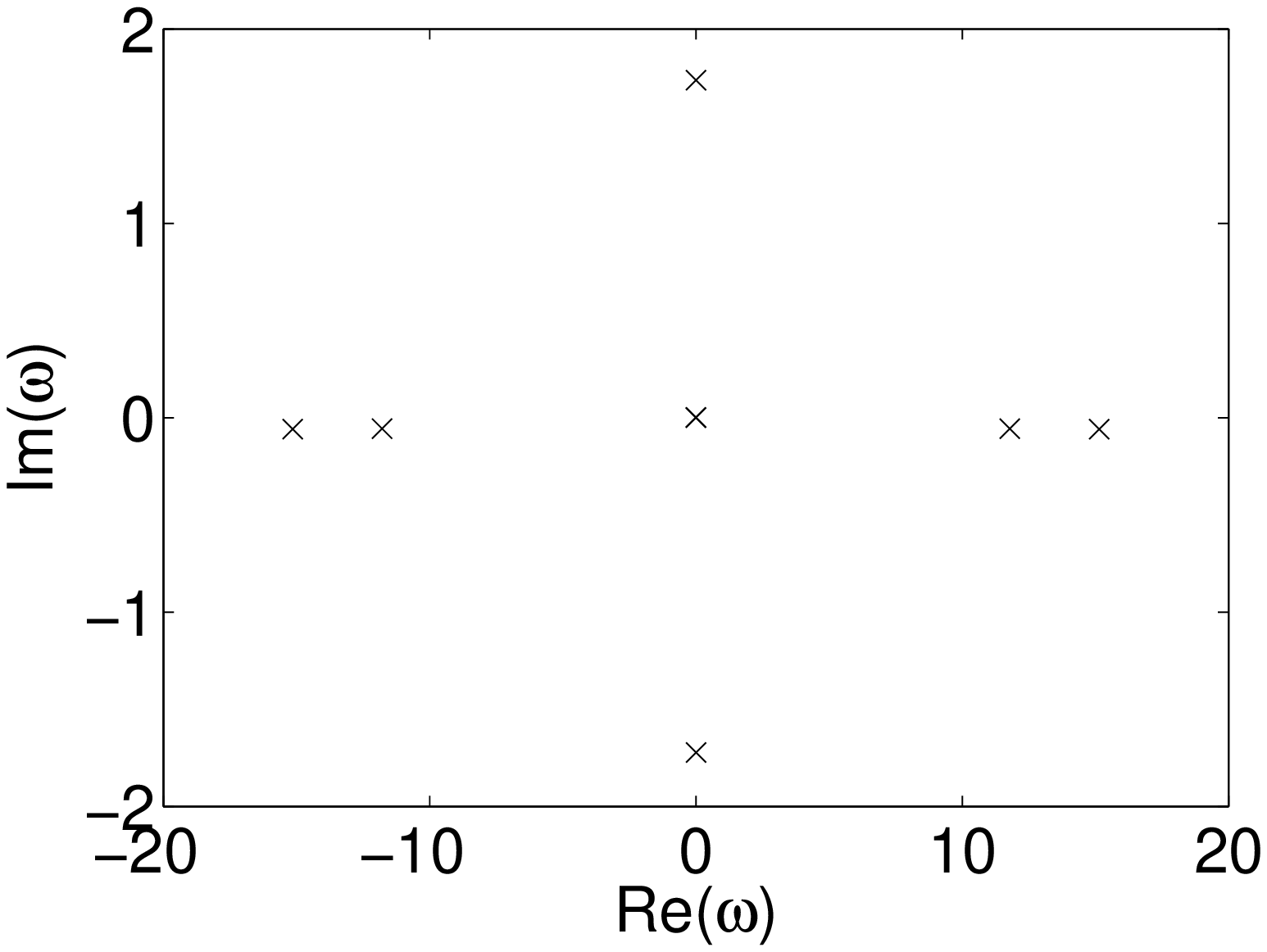}\\
\includegraphics[width=7cm,  angle=0]{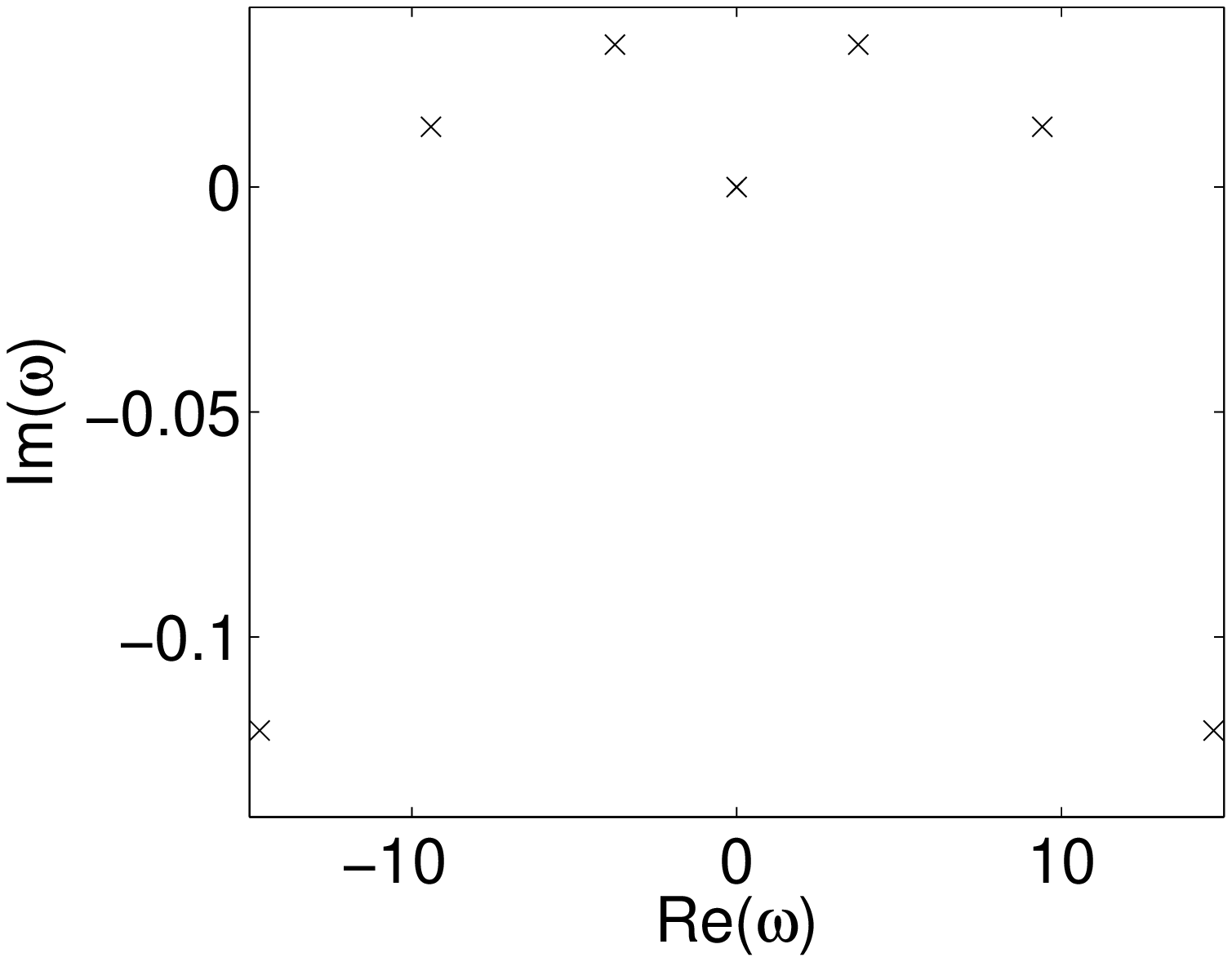}
\includegraphics[width=7cm,  angle=0]{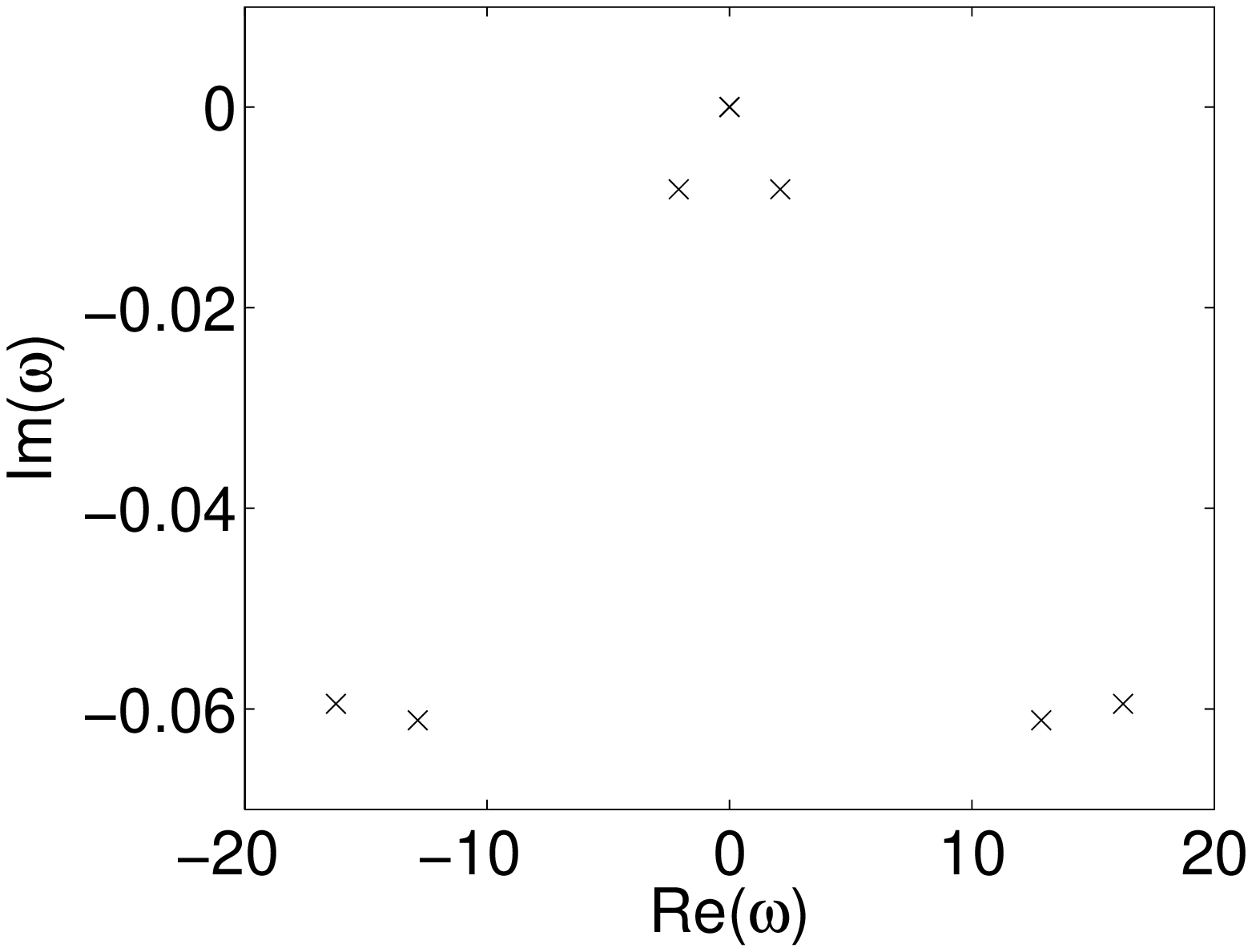}
\caption{\label{fig-DD_BdG} {Bogoliubov excitation spectra for the same parameters as in 
figure~\ref{fig-DD2} computed with $n_{\rm B}=4$ basis functions. Upper left: {autochtonous} self-trapping 
state (AuT), lower left: {allochtonous} self-trapping state (AlT), upper right: allochtonous almost 
antisymmetric state (Al-), lower right: allochtonous almost symmetric state (Au+)}} 
\end{figure}
Figure~\ref{fig-DD_BdG} shows the Bogoliubov excitation spectra for the same states as in
figure~\ref{fig-DD2} computed with $n_{\rm B}=4$ basis functions. Note that in the figure only seven
instead of eight eigenvalues are observed since the eigenvalue zero is doubly degenerate.
The results do not change significantly
if more basis states are taken into account. The excitation spectrum of the state Al- (upper right
panel) has an excitation with zero real part and a large positive imaginary part which leads to
instability as observed in the dynamical calculations (cf. figure~\ref{fig-DD_Dyn2}).
The autochtonous states (upper left and lower right panel) only have excitations with negative
imaginary part so that the adiabatic stability seen in the dynamical calculations (figures
\ref{fig-DD_Dyn1} and \ref{fig-DD_Dyn4}) is confirmed. The first (nonzero) excitation of the Au+ state  (lower right panel) has the eigenvalue $\omega \approx \pm 2.1 -\ri 0.0075$. The small imaginary part indicates that the excitation is only weakly damped. The period $2\pi/|{\rm Re}(\omega)| \approx 3$ of the excitation agrees with the period of the small oscillation in the background state's population imbalance (upper right panel of figure \ref{fig-DD_Dyn4}) for short times.
In the limit $g \rightarrow 0$ the nonzero eigenvalues of the BdG equations (\ref{DD-BdG}) are given by the differences of the eigenenergies of $H_0$ so that for longer times the oscillation period slowly approaches the value $2\pi \hbar/\Delta\mu_{g=0}\approx 7.53$ for the Josephson oscillations of the linear system.

The situation is more involved for the excitation spectrum of the allochtonous self-trapping
state (AlT) (lower left panel). In the dynamical calculation it seems to evolve adiabatically
until the bifurcation point is reached (figure~\ref{fig-DD_Dyn3}). Yet, its excitation spectrum
shows eigenvalues with positive imaginary parts. These imaginary parts
${\rm Im}(\omega)\approx 0.0317$ are, however, quite small and the characteristic time scale for
the growth of the respective excitations can be roughly estimated as $1/{\rm Im}(\omega) \approx 32$   
which is on the same order of magnitude as the time $\Delta t \approx 30$ that it takes for the dynamical
evolution to reach the bifurcation point (cf.~figure~\ref{fig-DD_Dyn3}). Consequently, the evolution
of the eigenstate is approximately adiabatic up to the bifurcation point if the initial population 
of the destabilizing excitations is small.

\section{Conclusion}
In this paper we analyzed the 
resonance states and the decay dynamics of the nonlinear Schr\"odinger
equation for a double delta-shell potential.

By means of an approximation the resonance wavefunctions and eigenvalues were calculated analytically.
In analogy to the the respective closed system the real parts of the eigenvalues in dependence on the nonlinearity $g$ show a saddle node bifurcation as symmetry breaking solutions emerge for a critical value of $g$. The imaginary parts (decay rates) undergo a similar bifurcation scenario. The approximate analytical results are in good agreement with numerically exact calculations based on complex absorbing potentials.

Comparison with a finite basis approximation demonstrates that both left and right eigenvectors of the linear $(g=0)$ system must be taken into account to obtain correct values for the decay coefficients with
a small number of (e.g.~two) basis functions.

A time--propagation of the eigenstates reveals that the states with linear counterpart (autochtonous states AuT and Au+) evolve according to an adiabatic approximation based on the quasisitationary resonance states for different values of the effective interaction $g \int_0^a|\psi(x)|^2 \rd x$. The time-evolution of the allochtonous self-trapping state (AlT) can also be described by the adiabatic approximation until  the bifurcation point is reached and the wavefunction starts to oscillate between the wells. For the allochtonous almost antisymmetric state (Al-) the wavefunction starts to oscillate immediately so that this state never evolves adiabatically. These results indicate that the self-trapping is not immediately destroyed by decay but is preserved on a time-scale determined by the decay coefficients of the respective self-trapping states. Furthermore it was shown that these adiabatic stability properties can also be deduced from a linear stability analysis based on the Bogoliubov-de-Gennes equations.


\ack
We thank James R.~Anglin for valuable discussions. Financial support via the DFG Graduiertenkolleg
792 "Nichtlineare Optik und Ultrakurzzeitphysik" is gratefully acknowledged.

\begin{appendix}
\section{Resonance solutions of the linear Schr\"odinger equation with an open double well potential}
\label{app_DDShell_lin}
The Galerkin-type approach in section \ref{subsec-DD_Galerkin} requires the computation of the resonance eigenfunctions $u(x)$ and corresponding eigenvalues $\mu-\ri \Gamma/2$ of the Hamiltonian $H_0$ given in equation (\ref{H0_u}) with the potential $V(x)$ given in equation (\ref{DDShell_pot}) which are obtained by solving the stationary Schr\"odinger equation
\be
  \left(-\frac{\hbar^2}{2m} \partial_x^2 +V(x)\, \right)u(x)=(\mu-\ri \Gamma/2)u(x)
\ee
with Siegert boundary conditions.
The ansatz
\be
 u(x)= \left\{
                    \begin{array}{cl}
                      \sin(kx)   & 0 \le x \le b \\
                      \left(\sin(kb)/\sin(kb+\vartheta) \right) \sin(k x+\vartheta)   & b < x \le a \\
                     \left(\sin(kb)/\sin(kb+\vartheta) \right)\sin(ka+\vartheta) \, \re^{{\rm i}k(x-a)}            &     x>a
                    \end{array}
              \right.
\ee
with $k=\sqrt{2m(\mu-\ri\Gamma/2)}/\hbar$ already makes the wave function continuous at $x=a,b$ and satisfies the Siegert boundary condition $\lim_{x \rightarrow \infty} u'(x)=\ri k \, u(x)$ as well as the boundary condition $u(0)=0$.
The matching conditions for the derivatives at $x=a,b$ read
\be
   k  \cos(ka+\vartheta)=(\ri k-2\lambda_a)\sin(k a+\vartheta)
   \label{DD_app_BC1}
\ee
and
\be
   k \cos(kb)=k\frac{\sin(kb)}{\sin(kb+\vartheta)}\, \cos(k b+\vartheta)- 2\lambda_b \sin(k b) \, ,
   \label{DD_app_BC2}
\ee
respectively. The complex quantities $k$ and $\vartheta$ are obtained by solving these equations numerically.
The wave function $u(x)$ diverges for $x\rightarrow \infty$ since ${\rm Im}(k)<0$. Therefore we use exterior complex scaling (see e.g.~\cite{Mois98,Zavi04}) to make the wave function square integrable.
The $x$ coordinate is rotated by an angle $\theta_c$ from the point where the potential $V(x)$ becomes zero. In our case this reads
\be
x \rightarrow \left\{
                    \begin{array}{cl}
                      x   & x \le a \\
         a+(x-a)\exp(\ri \theta_c)          &     x>a
                    \end{array}
              \right. \, .
\ee
In the scaled region the Schr\"odinger equation becomes $\exp(2\ri \theta_c)u''(x)+k^2 u(x)=0$. The equations (\ref{DD_app_BC1}) and (\ref{DD_app_BC2}) remain unaltered. For a sufficiently large rotation angle $\theta_c$ the wavefunction $u(x)$ becomes square integrable in $0 \le x < \infty$.

Because of $H_0=H_0^T$ complex conjugation of equation (\ref{DDShell_H0Dagger}) shows that the corresponding left eigenfunctions $v(x)$ are given by $v(x)=(u(x))^*$. We normalize the eigenstates such that
$\int_0^\infty \rd x \, v^*(x) u(x) =1$.
\end{appendix}

\section*{References}

\end{document}